# Part 1. During Inflation: Preferences, Investment, Risk*

**Leonid A. Shapiro**

**Abstract.** Producers cannot avoid being misled in their decisions even when they know inflation is taking place; effects of inflation are neither proportional to its extent nor to its magnitude — they resemble an infinite sum of disorienting impulses from perspective of entrepreneurs — and so slight inflation can, too, like greater inflation, force relations of market prices and profits to deviate significantly from such relations which would exist if intervention was indeed absent. This causes manufacture of products contrary to actual preferences of consumers and, too, real availability of capital goods. Harmful effects of intervention are themselves not linear and so minimizing it without eliminating it does not guarantee that harm arising from this intervention is minimal and thus analysis suggests that during every burst of inflation — slight or great — possibility of significant decrease of welfare cannot itself be rejected *a priori*.

## §1
## Introduction

Is increase of roundaboutness of production or lengthening of investment chains — and corresponding necessary decrease of quantity of first-order goods produced or manufacture of different first-order goods — proportional to increase of quantity of money or its substitutes, and supply of loanable funds, by way of credit expansion or coercive intervention, namely, inflation? Following Cantillon (1755), Craig (1821), Say (1821), Gossen (1854), Menger (1871; 1888), Hayek (1931; 1932), Hulsmann (1998), Hutt (1939; 1956), Mises (1912; 1949), Rothbard (1962a; 1962b), Garrison (2001):

1. Producers cannot avoid involvement in business cycles when inflation occurs. This remains true when they know that inflation is taking place and what it causes. Producers cannot distinguish changes in money-prices of economic goods caused by changes of preferences of consumers from such changes that originate with some individuals receiving additional money or fiduciary media by way of coercive intervention prior other individuals and spending it according to their own preferences while money and money-substitutes acquired and held by individuals receiving it later in market exchanges become less scarce, less valuable, and purchase fewer economic goods than prior inflation.

2. Producers can neither observe all exchanges taking place in society nor observe preferences of consumers directly. They learn preferences of consumers exclusively through observing changes in money-prices of economic goods, which are ranked among first-order goods in preferences as quantities of money-held. Market rate of interest everywhere indexes personal rates of interest together with risk. They have no way of distinguishing changes in market rates of interests caused by additional loanable funds originating with additional saving or decreases in time-preferences of consumers from changes in market rates of interests caused that are caused by additional loanable funds originating with additional credit or inflation. Ratio of revenue to quantity of their productive assets in terms of money — capital — is changed by inflation by way of changing distribution of purchases of various consumption goods. Money-prices of various employments of labor and other capital goods, which are derived from their contribution to changed revenue, are ultimately changed in this way, too. This causes factors of production to be employed in different ways and frequencies of production of various productive assets change.

Producers lengthen some investment chains in order to manufacture greater quantities of economic goods whose market prices had increased and shorten other investment chains. Decreased market rate of interest causes greater lengthening of some production structures than shortening of production structures that are not co-possible. Those individuals or industries who receive money earlier than other individuals or industries gain at their expense, but ultimately all groups of actors make decisions regarding work which they do which are not sustainable: they satisfy preferences of consumers less than other decisions they would have made if inflation had not taken place. Money-prices of alternative options are used by producers to determine what is more valuable to consumers and therefore how much of what is to be produced, and when, during comparisons of pairs of possible actions that are not co-possible. They respond to changes in money-prices and market interest rate by changing their investment chains. During inflation time-preferences did not change, and lengthening of investment chains or structures of production in response to lower market rates of interest and resulting change in direction of production, of what economic goods are produced instead of those economic goods that would have been produced otherwise, causes divergence of production from preferences that it was intended to satisfy.

3. Sufficient saving for such lengthening of investment chains to take place is ultimately non-existent, and many processes of production cannot be completed to deliver first-order goods from which value of higher-order goods involved in production is derived. Production diverges from satisfaction of actual preferences of consumers. Resources are withdrawn from producers that do satisfy actual preferences of consumers; they are outbid. Some economic goods that are less preferred by consumers are produced in greater quantities than other economic goods that more preferred by consumers. These cannot be sold in quantities that were anticipated by producers and yield losses of revenue. Capital is sum of time-discounted market value of assets, which are higher-order goods — indexed by money-prices — that are owned by producers. Significant quantities of higher-order goods have been wasted. Many higher-order goods that are specific to uses deep within investment chains, after investment chains are shortened, are useless. They are neither sold nor used because they lack value. Decrease in supply of capital, partly from physical waste of higher-order goods, leaving fewer of them, partly from loss of value during a crisis of many higher-order goods, consumption of capital, in other words, causes production of first-order goods and higher-order goods that is possible during the near future and the distant future to be less than it otherwise would have been without inflation. Many opportunities must be forgone that otherwise would not have necessarily been impossible.

4. Debt cannot result in future capital goods being consumed now. From perspective of present consumers, such capital goods do not exist in the present, but exclusively in the future. What does not exist presently cannot be consumed: that is physically impossible. Present capital goods are really consumed instead. This is done by politically-direct extra-market expansion of credit or legal tender resulting in re-distribution of relative quantities of money, money-substitutes, or fiduciary media, without consent of consumers that hold concrete these objects in the present, which shifts resources to those projects that are contrary to their preferences in the present and in the future. Such projects cannot be completed for lack of capital goods due to absence of change in consumer time-preferences, which is the part of income spent on first-order goods in the present relative to the part of income saved and invested, spent on production, in the present, or in other words, relative to the part of income spent on first-order goods in the distant future.

5. Lower market interest rate during expansion of quantity of money or money-substitutes causes investors to save and invest less of their income and instead consume more of it, because return to saving and investment is less in the present; but then consumption of capital in several directions is accompanied by simultaneous decrease in capital accumulation, and not only do time-preferences of consumers remain identical during inflation, so that lengthening of investment chains is contrary to time-preferences of consumers, but afterwards they actually change in direction contrary to lengthening of investment chains. This results in them satisfying consumer preferences even less. Distant future income, due to decrease in capital accumulation accompanied by consumption of capital, is significantly less than it otherwise would have been, which implies that fewer goods can be consumed in the distant future, while saving in the distant future is also necessarily less than it otherwise would have been. Consequences of inflation are long-term decrease in want-satisfaction of consumers by decreasing their future ability to produce economic goods.

Answering questions of proportion requires analysis of form of people's responses to arbitrary stimuli, namely, their preferences. All change of state, for instance, production, necessarily takes place over time. Every action, e.g., production, is multiply anticipatory (Rosen 1985a,b), doubly speculative (Mises 1949); it is anticipated to satisfy anticipated wants, because these wants are fundamentally in the future, and not certainly known in the present.

## §2, 3, 4

## Notation, Procedures, Corollaries

[Refer in this pre-print to the corresponding section in my paper: « Leibniz's Law of Continuous Change and Analysis of Fundamentally Uncertain Phenomena in Commerce. »]

[These two sections are discussions of the ideas of Fisher, Pareto, Cuhel, Wicksteed, Mises, and Bernardelli on the subject of preferences. Their works are cited below. A category theory model of preferences that generalizes their arguments is produced using methods in my paper: « Leibniz's Law of Continuous Change and Analysis of Fundamentally Uncertain Phenomena in Commerce. » The final analytical model is supplied there and some of that material is repeated. In this section more literature was reviewed but some mathematics from that paper was omitted. In one sentence: preferences are non-analytic distributions generated by generalized functions. Demand of consumers is not known but anticipated from perspective of managers and entrepreneurs. Relation of this to management of investments, subject of this paper, follows in the next section.]

## §5

## Analysis

Marginal quantities are concrete changes in state of environment at distinguishable moments of time, i.e., differently dated quantities of otherwise identical goods correspond to different numerical labels. Temporal preferences of consumers determines instantaneous personal rate of interest $\rho$; time-preferences, like other preferences, can of course be different, but one set of preferences cannot be *per se* greater or less than some other set of preferences; they consist of discrete taking or leaving, namely, choices (Mises 1949; Barnett & Block 2011).

From perspective of a person at a concrete moment, these are manifested by his or her splitting of all labeled marginal quantities into two sets each having infinitely many elements: first-order goods and higher-order goods. Earlier capital goods — present higher-order goods — can ultimately be treated as time-discounted future consumption — later first-order goods — which they are used to produce. Instantaneous split of momentary income into an ensemble of first-order goods which are consumed during the immediate future and remainder portion that is invested or saved — changed into higher-order goods if not already in this form — in order to produce future income over time, or in other words, into consumption and capital, if these sets of marginal quantities are indexed by sums of their present money-prices, results. (Craig 1821; Mises 1912; 1949; Rothbard 1962b)

Identical things at different times are different marginal quantities whose values are distinguishable. This implies that $\mathscr{R}\,\xi = \mathscr{F}\,(\xi\,;\rho)$ . *Observe that* $\rho$ *is one of the main parameters of all impulses that determines* $F\xi$ *by infinite summation*. We don't have to know what preferences are precisely in order to know that, invariant to change of preferences, substituting one $\rho$ in place of some other $\rho$ may possibly result in a great change of $F\xi$, e.g., entire re-arrangement of value hierarchies. Suppose inflation takes place; what happens then?

Producers are misled by inflation, by way of politically-determined extra-market expansion of credit or coercive intervention that directly increases aggregate supply of money, money-substitutes, or fiduciary media, which causes market interest rate to decrease; and why? Producers must anticipate future preferences of consumers, which involves determining time-preferences which they have; but this can be done solely by observing a market interest rate, and so their judgments during inflation are distorted. They anticipate entirely different consumer preferences; but they cannot do better: direct observation of time-preference isn't possible.

Producers infer present time-preferences of consumers — which fundamentally enter into determining all valuations that consumers make — from market interest rate. They speculate by inference from market rate of interest what are future preferences of consumers. Observe that changing parameters which appear in every single entry of some infinite series of impulses has potential to greatly change what result that series converges toward. Impulses are non-linear. In a hierarchical order that is entirely relative, like in systems of monads, every choice is a mirror of every other choice. Many slight changes of order are identical to complete re-ordering. Imagine a list that must be entirely re-organized, because every other element has changed places in that order. And so, producers must anticipate entirely different preferences from what consumers shall actually have in the future. This shall cause them to make a large number of choices regarding investment chains or what to produce that are contrary to actual consumer preferences in the future. Of course, later, they shall be very surprised to find such a great divergence between what they anticipated consumer preferences where and what they actually turn out to be. Let us assume that inflation is very great in extent. Anticipations of producers as to what preferences they must try to satisfy will diverge substantially from what these preferences actually are. What if inflation was slight in extent?

When slightly changing very many choices, the overall result may also be a great change. There is no *a priori* reason to suspect that changing the parameters of a very large, possibly infinite, series of impulses, every one different from every other, would result in linear change in overall result; but there is, formally, always great susceptibility for slight changes in many choices, to lead to something very different from what otherwise would result. It would often cause significant changes to the structure of production contrary to actual preferences of consumers, resulting in losses some time later, and contributing part of what define the future situation as a crisis. Damage to welfare, defined as mutual gain arising from voluntary exchanges (Bonnot de Condillac 1776; Destutt de Tracy 1817; Mises 1949; Rothbard 1956), would be substantial in both cases, because production shall be misled and pushed by inflation in a direction contrary to the values of consumers, reducing the number of elements in the set of possible mutually beneficial exchanges.

When the market rate of interest changes, anticipated preferences of consumers that producers use to determine the direction of investment shall change greatly, whether inflation was slight or great, and production shall be re-directed, not only contrary to actual time-preferences of consumers, but contrary to most other actual consumer preferences also.

Refer to the following figures.

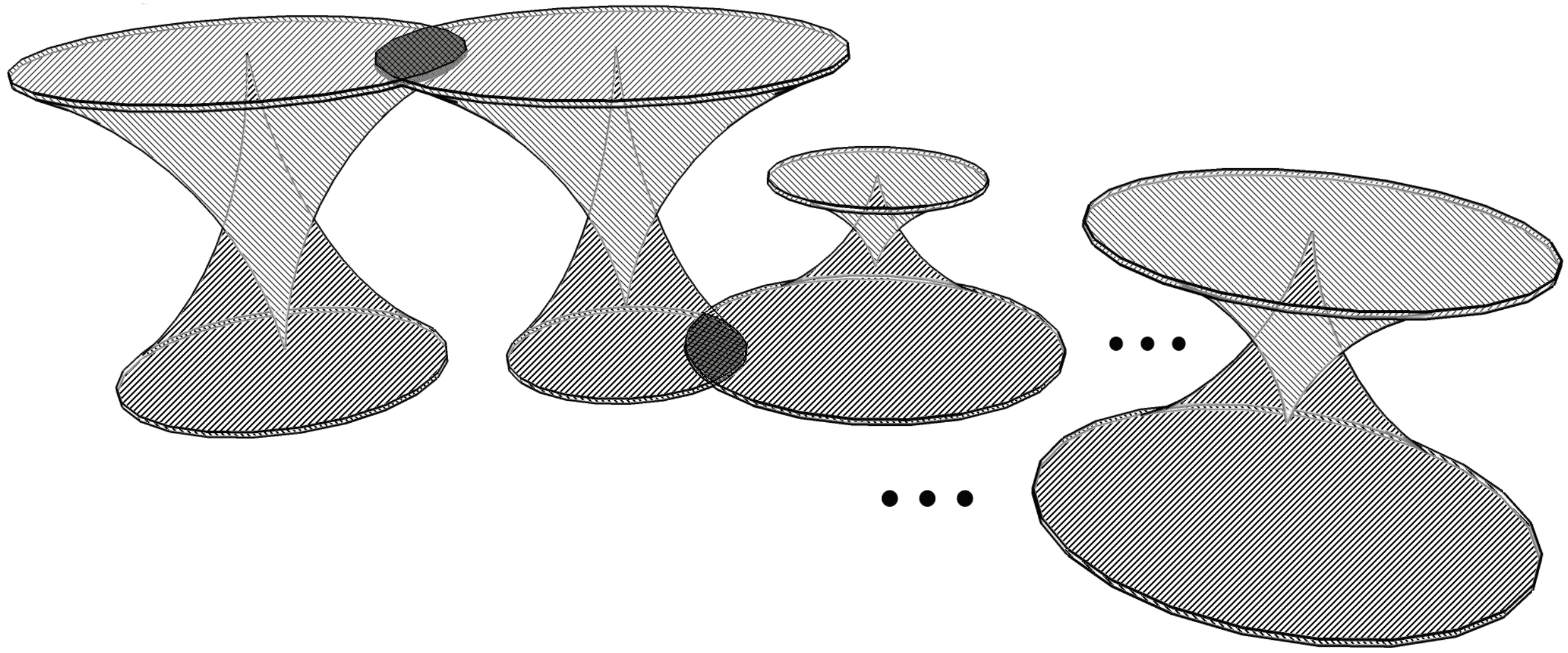

*Figure 3*. Whole market from perspective of its parts.

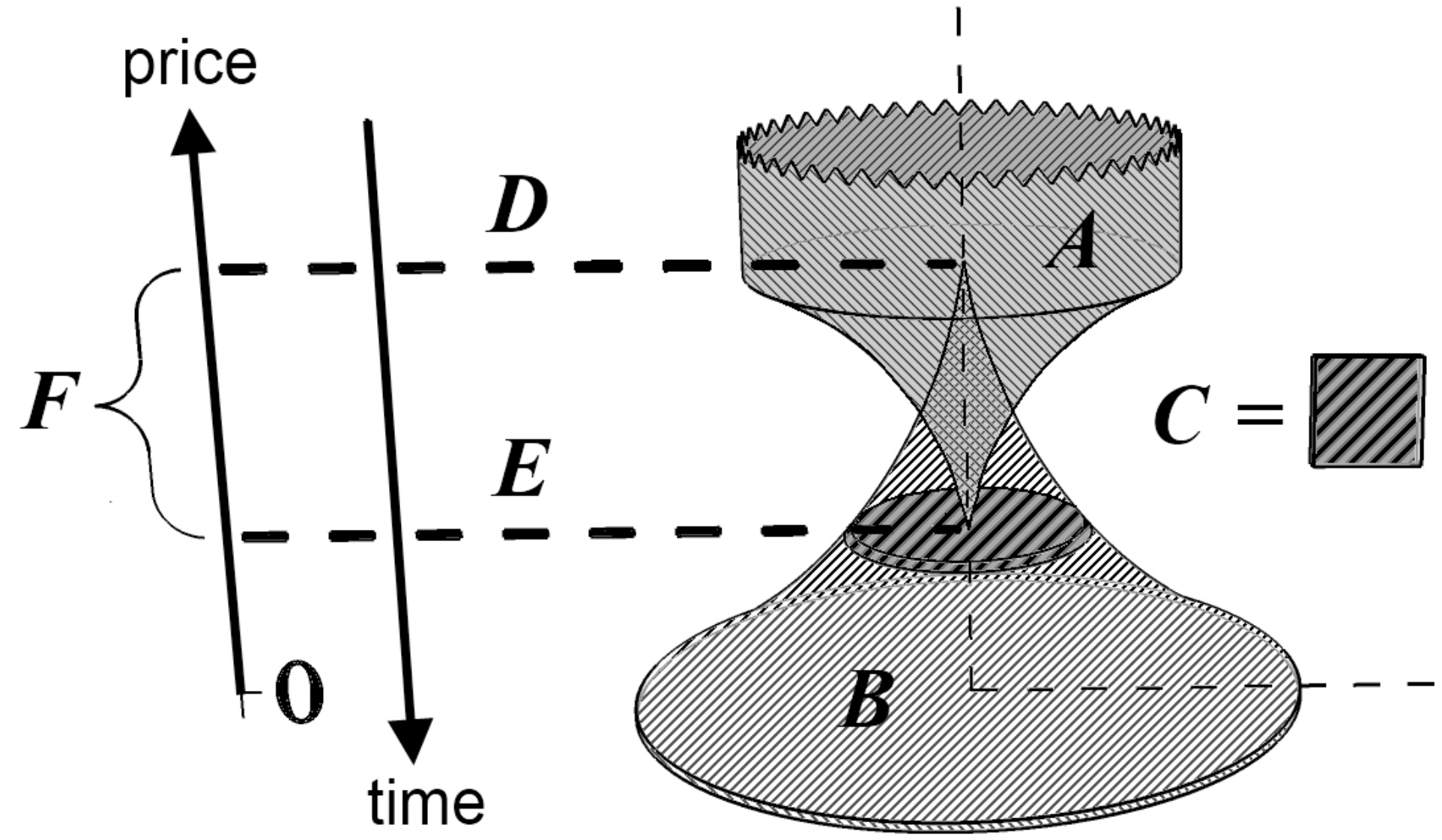

*Figure 4*. Analysis of a market.

This is a disaggregated price-volume stocks-flows model of one-to-one correspondence of marginal quantities bought and marginal quantities sold after such ensembles of goods were already manufactured by producers who began their production earlier in anticipation of future preferences of consumers from their perspective at that moment. Prices are quantities of goods which enter into exchange (Bonnot de Condillac 1776; Menger 1871), and a certain range of prices are in fact realized, and this interval (F) stretches from supremum (D) and infinitum (E). Following Craig (1821), imagine volumes (A) of different marginal quantities of various goods and volumes (B) of different buyers: A-type volumes of overlap if marginal quantities concerned are complements and B-type volumes overlap if they are substitutes at that price. These are distributions, too, neither sums nor volumes in anything except topological sense, namely, they are one-to-one mappings of price-time-slices (see next figure) of marginal quantities, to price-time-slices of their respective buyers, and so welfare is count-wise proportional to number of elements contained in overlaps of such pairs of volumes: these are mutually-beneficial voluntary exchanges.

What happens when producers anticipate preferences that do not truly belong to consumers and they re-direct production along those erroneous lines because they are misled by inflation or other change in market interest rate caused by coercive intervention? That situation resembles the first case in the next figure, where there is almost no overlap: many goods that are more abundant than buyers and few of them are sold, which gives the impression to some people on the street that there is a so-called general glut, superabundance. That is not the whole picture and physically impossible: people ignore the simultaneous shortages of other goods that were produced in quantities far less than could be sold — which necessarily occur by way of conservation of capital goods according to their scarcity causing many have been withdrawn to different production processes — and these consumers value more (Say 1803; 1814; 1821).

This is the second case in the following figure. Notice, also, that such bubbles are no longer connected with other markets by feedback relations. These cases existing side by side, disconnected, characterizes crisis situations. This is what causes people to experience additional cost or regret of foregone want-satisfaction that was otherwise possible (Bonnot de Condillac 1754b).

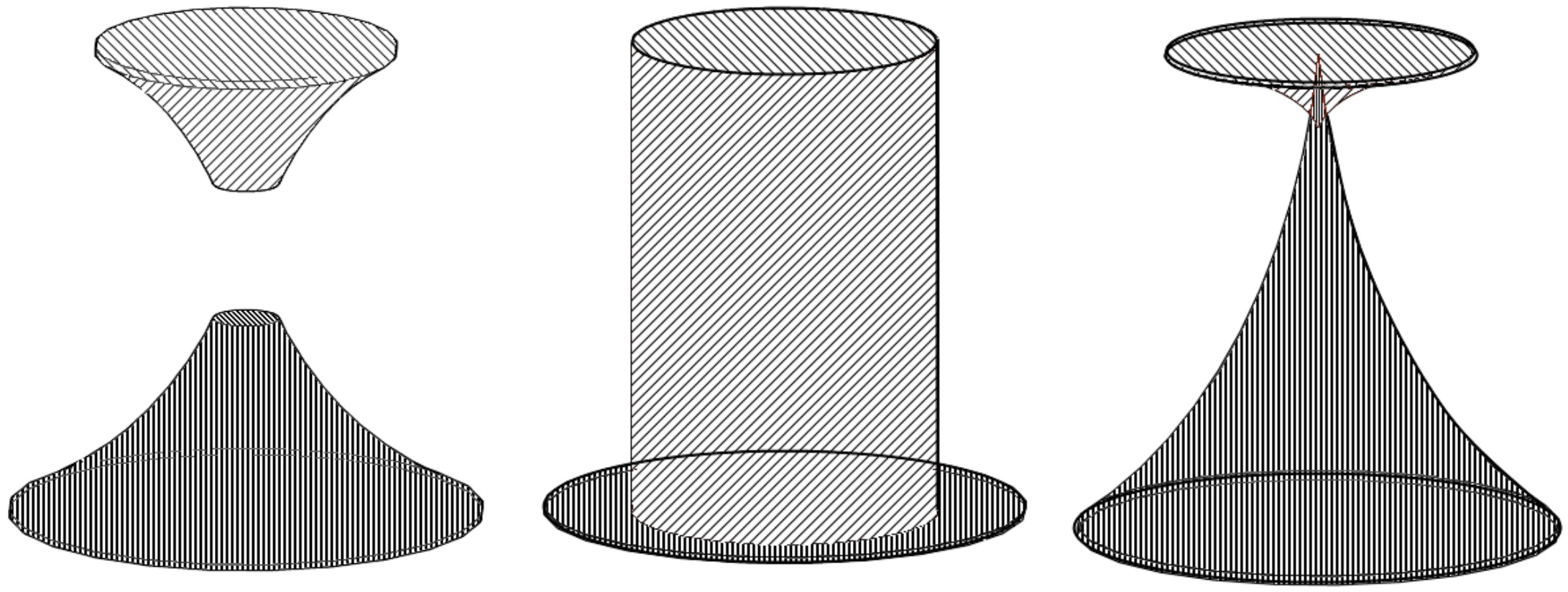

*Figure 5*. Crisis.

This is true for every coercive intervention that changes market interest rate. Such analysis might reveal other places where coercive intervention, both slight and great, may cause great harm by significantly disturbing market participants' anticipations of the future.

Frequency of « successful » events in a market, entrepreneurs to acting « successfully », i.e., satisfying preferences of consumers by truly anticipating them and producing without contradicting them, without « error », is possibly significantly reduced but does not converge to zero during inflation: « determining and anticipating choice are two different things; in order to *determine* some future event, we have to know all factors bringing it about; but we can *anticipate* human action without knowing any of these factors — we can even, for instance, anticipate it by chance — and for this reason alone successful action or equilibrium is possible under uncertainty.» (Hulsmann 2000)

Once money-indices correlated with consumer preferences — losses, market interest rate, and profits, for instance — are distorted by inflation, entrepreneurs are left without much information to guide their anticipations of preferences of consumers except chance guessing (Gossen 1854; Mises 1949). What can they do?

Entrepreneurs who lack true theoretical knowledge of effects of politically-determined extra-market credit expansion can obtain it; without such knowledge they cannot — except by guessing — anticipate effects of such credit expansion; indeed many decisions to lengthen investment chains or structures of production processes during credit expansion which later result in losses are made by a set of producers solely because they lack true theoretical knowledge of how credit expansion causes illusion of increasing real wealth when really increasing consumption and investment chains together without saving is consumption of capital goods and must result in failure to produce salable products or manufacture solely of products salable at prices consumers refuse to purchase them at once such expansion of supply of credit in their society ceases (Craig 1821; Mises 1949; Bagus 2007).

« If it is possible that effects of inflation are correctly anticipated then inflation does not necessarily lead to error » (Hulsmann 1998). Nothing necessarily prevents entrepreneurs from correctly anticipating effects of inflation: they might guess true effects of inflation correctly, or infer true effects of inflation correctly from false reasons, which is always possible (Gettier 1963; Hulsmann 1998; 2000).

Entrepreneurs cannot actually distinguish changes in market rate of interest originating with politically-determined extra-market expansion of supply of credit in their society by coercive intervention from those increases or decreases of it originating with increase in saving and investment of consumers, and this is especially true where such credit is made legal tender and its acceptance as payment for goods during buying and selling is enforced by coercive intervention too (Mises 1912; 1949). If they prefer not to cease production of goods during this « inflation », which they do or do not anticipate, then they cannot avoid taking and investing additional money or its substitutes introduced by coercive intervention somewhere in their society, too. Why? If they did not do so, then their competitors would actually do so and outbid them when purchasing factors of production and this would prevent them from continuing to manufacture, re-sell, or sell products, but they can speculatively judge that inflation is taking place and then they can possibly anticipate future prices of factors of production and products being greater than past prices of these marginal quantities and speculatively attempt to sell products of their investment chains at prices they judge are sufficiently increasing over time in order to generate money-revenue exceeding their money-costs which increase over time (Hulsmann 1998).

If some entrepreneurs have theoretical knowledge of effects of such expansion of supply of credit in their society, then such entrepreneurs they can make a speculative decision. Sometimes they can make a speculative judgment that inflation is taking place, and they estimate its extent and magnitude, too; but inflation arises from political coercive intervention in market processes; that fact alone — its arbitrariness — makes such figures not easy to gauge without great divergence of results from true measurements. Thus origin of inflation in coercive intervention — not inflation itself — causes a significant part of expansion of investment chains contrary to time-preferences of consumers during inflation and welfare-reducing future losses which result from this (Hulsmann 1998). Then, while anticipating future revenue from alternative manufacturing processes, entrepreneurs must suppose that time-preferences of consumers did not decrease, which implies that observed decrease in market interest rate was caused by inflation, and that caused in turn, from their perspective, proportional increase in riskiness of all manufacturing processes in that market. Such judgment can prevent some losses which these produces otherwise would ultimately suffer during crises, for instance, one or several manufacturing processes they would have begun or lengthened erroneously during inflation would not be started, because anticipated future money-revenue of these possible additional investments discounted by market interest rate and risk would not be erroneously anticipated to exceed money-costs of these possible additional investments. This speculative procedure cannot fully replace a market interest rate when that is not distorted by inflation: what is normally indexed with certainty by market interest rate — quantity of capital goods available in society and time-preferences of consumers — are by way of inflation made to become a source of additional fundamental uncertainty plus some additional stochastic risk.

In other words, inflation transforms a system which usually satisfies wants of its participants — where judgments regarding contracting, expanding, starting, and stopping manufacturing processes are objectively indexed by indices, losses, and profits, which are strictly coupled to such preferences, and speculative decisions are made using these observables as guides (Mises 1949; 1952; Hulsmann 2000) — into an exercise of guessing what hidden events take place with no or almost no information about their causes, because they are really arbitrary interventions, « fair » guessing, « gambling ». This removes part of natural coupling of production and wants of consumers of products so manufactured.

## §6

## Examples

## §7

# Part 2. Leibniz's Law of Continuous Change & Analysis of Fundamentally Uncertain Phenomena in Commerce

**Abstract.** Various uses of hypothesis of continuity of change of magnitudes in economies are precisely analyzed. Human actors always have wants which are not already satisfied and so action never ceases; it is result of discrete choices between pairs of possible but not co-possible different modes of behavior without intermediate degrees. Such choices are unique events. They are caused by preference which changes over time in an environment that itself changes over time — for instance, it is changed by such action, which takes place in it — and whose future configuration is not certain from perspective of actors whose environment it is. Such absence of certainty is fundamental to behavior that is partly determined by preference, which is to say, in other words, that change of their environment is one cause of it but not sole cause of it. This implies that profit and loss are themselves always possible in economies. Relation of this fact to opportunities of entrepreneurs to earn profit and estimation of risk — measurable absence of certainty — by managers of manufacturing processes are discussed.

## §1
## Introduction

What is truly continuous in economies? Leibniz's law of continuous change (LLCC) determines form of all human action or possible behavior which is actually done because somebody anticipates that it causes satisfaction of his or her anticipated future wants. This is a possible hypothesis and indeed several writers explicitly claimed that it is true (e.g., Marshall 1890) and during their analyses of such action other thinkers (e.g., Gossen 1854; Fisher 1892; Pareto 1906a) had tacitly assumed that it is true: but are these assumptions true? Many past writers who believed or inferred that these assumptions are true thought — also — that LLCC implies qualities of actors which it truly doesn't imply, for instance, that discrete categories of behavior — or discrete states of mind which cause it — don't exist and that all mental states and responses of bodies which these mental events cause merge one into other as distance from one increases and distance to other diminishes in that phase-space which associates experiences and responses in mind of each person (Marshall 1890), that marginal utilities or values or their differences are continuously varying quantities which can be added, subtracted, multiplied, and divided (Gossen 1854; Jevons 1871; Marshall 1890), that such quantities corresponding to infinitesimal parts of each person can be summed together over infinite continuous time and that resulting sums can themselves vary continuously (Edgeworth 1881), that monotone preferences which rank quantities of goods are continuous (Mas-Colell 1985), that order of consumption of concrete quantities of goods is arbitrary, or that preferred order of such consumption by way of continuous summation is determined separately from those preferences which order possible combinations of consumed marginal quantities whose value he or she feels is identical and not distinguishable (Pareto 1906b; 1909).

Many modern procedures — for instance, analyses presented by Allen (1934), Hicks (1934), and Mas-Colell (1985) — are themselves not substantially different from Fisher's (1892) procedures, whose analysis itself was not wholly free of all these past errors (Bernardelli 1934; 1938; 1939; 1952; 1954), and so these past errors are frequently repeated. Pareto (1906a,b) found again those procedures which Fisher (1892) found earlier; he omitted several true theorems which Fisher explicitly earlier presented and instead of them he presented several theorems which Fisher himself earlier omitted because they were not true. Pareto's (1906a) important contribution was his explicit presentation of those true propositions about value and welfare whose truth was implied but not explicitly proposed by Fisher (1892).

Many analyses which are derived from Fisher's work are less useful than his own work, because they actually consist of Fisher's procedures in combination with one or more false supplementary hypotheses or simplifying assumptions which are not necessary at all in order to generate phenomena which they are supposed to cause in models and they are introduced solely in order to insert preferences of authors of models into these models themselves. This fact is usually enthusiastically and proudly admitted by these authors themselves in those rare instances where they observe themselves doing this; but this is not science. If someone says that « apples are red and round and solid, because red apples are good », then really he or she altogether says nothing true about apples themselves; indeed apples are round and solid but some apples exist which are not red apples. This proposition tells those people who hear it nothing more than that he or she who said it likes red apples; but science is work done in order to learn true causes of phenomena and it is not work in order to learn arbitrary or conditioned preferences of observers of phenomena and how they prefer — like or dislike — various phenomena which are observed and their possible or true causes. Learning periodically over time facts or theorems which he or she did not already know is each person's own source of possibly endless fun observed Gossen (1854) and so authors of such models have an opportunity to learn with pleasure from Newton (1687) that those people who truly work in order to find truth « admit no more causes of natural things than those causes which are true and sufficient in order to explain their appearances, because Nature does nothing in vain and more is in vain there where less is sufficient. Nature lacks pomp of superfluous causes. »

For instance: want-satisfying behavior — which consists of a person taking or leaving or working to obtain one good instead of working to obtain another good — when several goods which are possibly taken or obtainable are not co-possibly taken or not co-obtainable goods is already wholly generated by association of hierarchically ordered labels with replacements of one space-time event by another space-time event without quantitative marginal utilities and values being associated with these replacements. These quantities were thus excluded from set of causes of conscious action of persons in models of such action once scientists learned that quantitative marginal utility and value are therefore superfluous causes of instances of all such purposeful behavior (Fisher 1892; Cuhel 1907; Mises 1912; 1949).

What is relation of continuous models and discontinuous models? Many writers (e.g., Fisher 1892) believed that discontinuous models « modify or extend » continuous models, which implies that they believed that purposeful behavior can be generated by continuous models and therefore that such systems and discontinuous models are homologous, alternative descriptions, « paraphrases », which are realized by identical natural systems (Rosen 1975). A question can be asked here: continuous models of economies and their discontinuous models are really alternative descriptions or not alternative descriptions?

Behavior which is done because somebody anticipates that it causes satisfaction of his or her anticipated future wants and those wants themselves cannot vary by way of infinitely small increments, because he or she cannot possibly realize infinitesimal movement and infinitely small perceptions are perceived but they are not consciously perceived (Leibniz 1896; Dosch & Sieroka 2008). Infinitesimal perceptions therefore neither satisfy wants nor cause such action which is anticipated to satisfy wants and such action and magnitudes which determine it must therefore vary discretely and indices of value of marginal quantities cannot vary continuously — they cannot vary by way of infinitely small differences (Rothbard 1956). All models of such action in economies and valuations which cause it where magnitudes that determine it and magnitudes of such action itself vary continuously imply contrary to hypothesis identity of things which are distinguishable by hypothesis, which is self-contradiction. All such action and valuations which cause it are discrete.

This is so; but discrete action leaves opportunities to earn profit by way of arbitrage — claimed several writers (e.g. Stigler 1952) — which is to say, in other words, that sellers who break apart lumpy quantities can sell more goods of a stock than sellers who do not break apart these quantities, because they can offer smaller quantities at lower prices to buyers who desire smaller quantities, while sellers who sell lumps of greater quantities of identical goods cannot sell to such buyers, and that in turn causes their profit to be greater than sellers who sell lumps of greater quantities of identical goods where cost of break apart lumpy quantities is less than additional revenue caused by doing so. Thus people do so, profit, get rid of « lumpiness of quantities », and cause magnitudes in economies to vary continuously in the limit. These writers then assume continuous variation in the limit of economic magnitudes, but that is false. Decrease of lumpiness being profitable does not imply that limits to reducing lumpiness of behavior do not exist. Infinitesimal quantities are only perceived unconsciously and so cannot determine action of consumers. All human action is taking or leaving and fundamentally discrete.

There is a larger problem with deterministic continuous models as they are generally used. All such models of action in economies and valuations which cause it which consist of related functions necessarily cannot generate uncertainty observed in real life; but this uncertainty is what causes profit or loss by way of actors purchasing goods at certain prices and later selling them — or products which result of their combination at prices — which are not certain from perspective of these actors when they earlier purchased all those goods (Cantillon 1755; Turgot 1769; Knight 1921; Mises 1949).

What hypothesis is called LLCC? In nature, « omnis mutation fiat per gradus », which is to say that « every change takes place gradually » (Leibniz 1695), or in other words, « every change or motion is continuous and takes place by passing through all intermediate degrees without leaps ... If difference of one cause and another cause continuously decreases, becoming smaller and smaller in process of disappearing, so that one cause eventually disappears into that other cause, then difference of their effects itself also continuously decreases, too, becoming smaller and smaller in process of disappearing, and eventually it ceases to exist ... Nothing is accomplished all at once. We pass always from small things to great things and from great things to small things everywhere through intermediate things, in degrees as in parts, and great motion never springs at once from rest and motion is not reduced to rest except through becoming smaller motion. » (Leibniz 1687; 1860; 1896)

This implies that « each part of each whole is itself divisible into several smaller parts » (Bernoulli 1713). LLCC is not identical to hypothesis that all solutions of systems of differential equations are themselves analytic functions — which are called continuous-smooth functions, namely, functions which are differentiable infinitely-many times at each point in their domain — because that hypothesis is not true — non-analytic functions integrate such systems too — which implies that LLCC is true, a law of nature, if and only if it not identical to that hypothesis (Truesdell 1984). LLCC cannot imply that analytic functions integrate all systems of differential equations. Many solutions of systems of ordinary-differential equations or partial-differential equations are analytic maps which are not themselves functions (Sato 1958a,b; 1959; 1960). In this essay LLCC is interpreted in terms of such maps, which makes it true indeed.

This is shown, then, to generate uncertainty in economic phenomena independent of unpredicted environmental changes, which make arbitrage opportunities possible in principle when these are absent and therefore always possible sources of profit or loss. Finally, corollaries of this fact for manipulation of practical data generally encountered in business are developed.

## §2

## Notation

For conciseness, because composition appears here more frequently than multiplication, left-to-right-combination juxtaposition is composition but multiplication is explicitly written: $E \circ \iota F = E(\iota)F = E\iota \cdot F$ . Identity functor $J = \{ \iota \mid F\iota = F = \iota F \} = JJ = J^{/}$ ; it is that functor that is its own inverse $J^{/} = \{ \iota \mid J\iota = J = \iota J \}$. Thus $J \neq F \neq F^{/} \neq J$ . (Menger 1944; 1952; 1953)

For instance: a combination $(J \mapsto F)(R \in X) = (J(R \in X) \mapsto (F(R \in X) = ((FR = P) \in (FX \subseteq Y))) = (R \overset{F}{\mapsto} P) = (X \overset{F}{\rightarrow} Y)$ is part of a category $I \overset{E}{\rightarrow} X \overset{F}{\rightarrow} Y \overset{\Gamma}{\leftarrow} I$ that itself « commutes » if and only if $EFI = FX = \Gamma I$ (MacLane 1971). Let $(((AB)(EF))((GH)(XY)))Z = AB.EF : GH.XY \therefore Z$ (Peano 1889). There where this is necessary in context $A / B$ is really set $A$ excluding set $B$ instead of quantity $A$ divided by quantity $B$. Let dom $U$ = dom $V$ = dom $W \mid U = V \neq W \mid$ cod $U$ = cod $V$ = cod $W$ . Various types of labels, marks, and names exist. For instance: « quantitative variables » are results of measurement — ordered associations of observables and their states or maps of natural things to real numbers — but « variable quantities » are constant functions or other functions — maps of real numbers to real numbers — and this fact limits order of composition of maps: because real numbers cannot be injected into « quantitative variables » and natural things are not in domain of maps which are « variable quantities » and so they cannot be injected into them (Menger 1952; 1953). An ensemble is a set which consists of several elements. Thus all those events or natural things which are not themselves marked by one name $X = XY \cup Xy \cup (\varnothing = XY \cap Xy)$ are each one really marked by a name $x = \text{non-}X$ (DeMorgan 1860; Jevons 1864). Let $(J, _) = (J, J) = (_, J)$.

Real part $\Re = (J - \overline{J}\,)\,/\,2 = \log(\exp(J \cdot \pi))\,/\,\pi = \{ (J - (\iota \in \mathbb{R}) \cdot i) \in \mathbb{R} \}$. Imaginary part $\Im = \mathrm{op}\,(\overline{J} - J)\,/\,(2 \cdot i) = \log(\exp(J \cdot j))\,/\,\mathrm{op}\,(j\,/\,i)$ there where $\mathrm{op} = 0 - J$. Thus $\overline{J} = 2 \cdot \Re\, J + \mathrm{op}\, J$. Let symbol $i = \sqrt{\mathrm{op}\,1}$ and $2 \cdot \pi \cdot i = j$. If $(x, z) \in \mathbb{R} \times \mathbb{R}$, then $x + i \cdot z = y \in \mathbb{C}$.

Let $(\tau, \varpi, \sigma) \in \mathbb{R} \times \mathbb{R} \times (\mathbb{R}\,/\,\{0\})$: so that $(\tau - i \cdot \varpi, \tau + i \cdot \varpi) = (\kappa, \overline{\kappa})$. Thus $(\kappa + i \cdot \sigma, \kappa - i \cdot \sigma) \in \{ \iota \in \mathbb{C}\,/\,\kappa \} \times \{ \overline{\iota} \}$. If so, then « hyper-function » $J = \lim_{\sigma \to \Im(JJ^{\prime})+} \{ J_+ (JJ^{\prime} + i \cdot \sigma) - J_+ \overline{(JJ^{\prime} + i \cdot \sigma)} \} = \mathrm{hyp}\,(J_+, J_-)$. For example: $J\kappa = \mathrm{hyp}\,(J_+\kappa, J_-\kappa) = \mathrm{hyp}\,(J_+, J_-)\,\kappa$. (Kothe 1952; Sato 1958a,b; 1959; 1960)

If we so desire, then we can perceive discrete change and discrete events continuously. For instance:

$$(J^2)^{1/2} = \frac{2}{\pi} \cdot \int_0^{\infty} \frac{J^2 \cdot dx}{J^2 + x^2} = \frac{4}{\pi} \cdot \int_0^{J} \frac{J^2 \cdot dx}{J^2 + x^2} = \begin{cases} J & \text{if } J \geq 0 \\ \mathrm{op}\,J & \text{if } J < 0 \end{cases}$$

(Cauchy 1844). Thus

$$\mathrm{sgn} = (J^2)^{1/2}/J = \frac{(1 + \mathrm{sgn})}{2} - \frac{(1 - \mathrm{sgn})}{2}$$

$$= \begin{cases} 1 & \text{if } J > 0 \\ \mathrm{op}\,1 & \text{if } J < 0 \end{cases} = \begin{cases} 1 & \text{if } J > 0 \\ 0 & \text{if } J < 0 \end{cases} - \begin{cases} 0 & \text{if } J > 0 \\ 1 & \text{if } J < 0 \end{cases}$$

$$= \mathrm{hyp}\left( \frac{\mathrm{op}\,1}{j} \cdot \int_1^{\mathrm{op}\,J} \frac{dy}{y}, _ \right) - \mathrm{hyp}\left( \frac{1}{j} \cdot \int_1^{J} \frac{dy}{y}, _ \right)$$

$$= \mathrm{hyp}(\mathrm{op}\,\log\,\mathrm{op}\,/\,j, _) - \mathrm{hyp}(\log\,/\,j, _)$$ (Sato 1959).

Very similarly

$$\mathrm{op}\,\mathrm{sgn} = \frac{(1 - \mathrm{sgn})}{2} - \frac{(1 + \mathrm{sgn})}{2}$$

$$= \begin{cases} \mathrm{op}\,1 & \text{if } J > 0 \\ 1 & \text{if } J < 0 \end{cases} = \begin{cases} 0 & \text{if } J > 0 \\ 1 & \text{if } J < 0 \end{cases} - \begin{cases} 1 & \text{if } J > 0 \\ 0 & \text{if } J < 0 \end{cases}$$

$$= \mathrm{hyp}(\log\,/\,j, _) - \mathrm{hyp}(\mathrm{op}\,\log\,\mathrm{op}\,/\,j, _).$$

Let $\mathrm{nul} = \lim_{\upsilon \to 0+} . \; J(JJ' + \upsilon)$. If so, then:

$$\theta_{\mathbf{X}-\mathbf{R},1} = \begin{cases} 1 & \text{if } \mathbf{X} \geq \mathbf{R} \\ 0 & \text{if } \mathbf{X} < \mathbf{R} \end{cases} = \begin{cases} 1 & \text{if } \mathbf{X} - \mathbf{R} \geq 0 \\ 0 & \text{if } \mathbf{X} - \mathbf{R} < 0 \end{cases} = \prod_{V=1}^{T} \theta_{X_V - R_V, 1}$$

$$= \prod_{V=1}^{T} \frac{1 + \begin{cases} 1 & \text{if } X_V - R_V \geq 0 \\ \mathrm{op}\,1 & \text{if } X_V - R_V < 0 \end{cases}}{2} = \prod_{V=1}^{T} \frac{1 + \mathrm{nul}\ \mathrm{sgn}\,(X_V - R_V)}{2} ;$$

$$\theta_{\mathbf{X}-\mathbf{R},0} = \begin{cases} 0 & \text{if } \mathbf{X} \geq \mathbf{R} \\ 1 & \text{if } \mathbf{X} < \mathbf{R} \end{cases} = \begin{cases} 0 & \text{if } \mathbf{X} - \mathbf{R} \geq 0 \\ 1 & \text{if } \mathbf{X} - \mathbf{R} < 0 \end{cases} = \prod_{V=1}^{T} \theta_{X_V - R_V, 0}$$

$$= \prod_{V=1}^{T} \frac{1 - \begin{cases} 1 & \text{if } X_V - R_V \geq 0 \\ \mathrm{op}\,1 & \text{if } X_V - R_V < 0 \end{cases}}{2} = \prod_{V=1}^{T} \frac{1 - \mathrm{nul}\ \mathrm{sgn}\,(X_V - R_V)}{2} ;$$

$$\theta_{\mathbf{R}-\mathbf{X},1} = \begin{cases} 1 & \text{if } \mathbf{X} \leq \mathbf{R} \\ 0 & \text{if } \mathbf{X} > \mathbf{R} \end{cases} ; \; \theta_{\mathbf{R}-\mathbf{X},0} = \begin{cases} 0 & \text{if } \mathbf{X} \leq \mathbf{R} \\ 1 & \text{if } \mathbf{X} > \mathbf{R} \end{cases} ;$$

$$\delta_{\mathbf{R}}^{\mathbf{X},1} = \begin{cases} 1 & \text{if } \mathbf{X} = \mathbf{R} \\ 0 & \text{if } \mathbf{X} \neq \mathbf{R} \end{cases} = \prod_{V=1}^{T} \delta_{R_V}^{X_V} = \prod_{V=1}^{T} \begin{cases} 1 & \text{if } X_V = R_V \\ 0 & \text{if } X_V \neq R_V \end{cases} = \prod_{V=1}^{T} \begin{cases} 1 & \text{if } X_V = R_V \\ 0 & \text{if } X_V < R_V < X_V \end{cases}$$

$$= \theta_{\mathbf{X}-\mathbf{R},1} - (\theta_{\mathbf{X}-\mathbf{R},0} + \theta_{\mathbf{R}-\mathbf{X},0}) = \theta_{\mathbf{R}-\mathbf{X},1} - (\theta_{\mathbf{R}-\mathbf{X},0} + \theta_{\mathbf{X}-\mathbf{R},0})$$

$$= \theta_{\mathbf{X}-\mathbf{R},1} - (\vartheta_{\mathbf{X}-\mathbf{R},1} + \vartheta_{\mathbf{R}-\mathbf{X},1}) = \theta_{\mathbf{R}-\mathbf{X},1} - (\vartheta_{\mathbf{R}-\mathbf{X},0} + \vartheta_{\mathbf{X}-\mathbf{R},0}) ;$$

$$\delta_{\mathbf{R}}^{\mathbf{X},0} = \begin{cases} 0 & \text{if } \mathbf{X} = \mathbf{R} \\ 1 & \text{if } \mathbf{X} \neq \mathbf{R} \end{cases} = \theta_{\delta_{\mathbf{R}}^{\mathbf{X},1} - 1, 0} = \theta_{1 - \delta_{\mathbf{R}}^{\mathbf{X},1}, 0} ;$$

$$\theta_{\mathbf{X}-\mathbf{R},1} \cdot \theta_{\mathbf{Y}-\mathbf{P},1} = \begin{cases} 1 & \text{if } \mathbf{X} \geq \mathbf{R} \ .\cap\!: \ \mathbf{Y} \geq \mathbf{P} \cdot \\ 0 & \text{if } \mathbf{X} < \mathbf{R} \ .\cup\!: \ \mathbf{Y} < \mathbf{P} \cdot \end{cases} ;$$

$$\theta_{\mathbf{X}-\mathbf{R},1} \cdot \theta_{\mathbf{Y}-\mathbf{P},0} = \begin{cases} 1 & \text{if } \mathbf{X} \geq \mathbf{R} \ .\cap\!: \ \mathbf{Y} < \mathbf{P} \cdot \\ 0 & \text{if } \mathbf{X} < \mathbf{R} \ .\cup\!: \ \mathbf{Y} \geq \mathbf{P} \cdot \end{cases} ;$$

$$\theta_{\mathbf{X}-\mathbf{R},0} \cdot \theta_{\mathbf{Y}-\mathbf{P},0} = \begin{cases} 0 & \text{if } \mathbf{X} \geq \mathbf{R} \ .\cap\!: \ \mathbf{Y} \geq \mathbf{P} \cdot \\ 1 & \text{if } \mathbf{X} < \mathbf{R} \ .\cup\!: \ \mathbf{Y} < \mathbf{P} \cdot \end{cases} .$$

Let $\mathbf{X} \subseteq \mathbf{Y} \mid \mathbf{R} \subseteq \mathbf{P}$ . For instance: this is true in events where $\mathbf{R} < \mathbf{P} < \mathbf{X} = \mathbf{Y}$ .

If also $\mathbf{X} = \sum_{V=1}^{T} .\, X_V \cdot \boldsymbol{\varepsilon}^V \Big|_{X_V \le Y_V} \in \mathop{\times}_{V=1}^{T} \mathbb{R} \Big|_{T \le I}$ and $\mathbf{Y} = \sum_{V=1}^{I} .\, Y_V \cdot \boldsymbol{\varepsilon}^V \in \mathop{\times}_{V=1}^{I} \mathbb{R}$ ,

$$\text{then:} \quad \vartheta_{\mathbf{X}-\mathbf{R} \mid \mathbf{P}-\mathbf{Y},1} = \begin{cases} 1 \text{ if } \mathbf{R} < \mathbf{X} .\cap\!: \mathbf{P} < \mathbf{Y} . \\ 0 \text{ if } \mathbf{R} > \mathbf{X} .\cup\!: \mathbf{P} > \mathbf{Y} . \end{cases} = \begin{cases} 1 \text{ if } \mathbf{X}-\mathbf{R} > \mathbf{0} .\cap\!: \mathbf{Y}-\mathbf{P} > \mathbf{0} \cdot \\ 0 \text{ if } \mathbf{X}-\mathbf{R} < \mathbf{0} .\cup\!: \mathbf{Y}-\mathbf{P} < \mathbf{0} \cdot \end{cases}$$

$$= \begin{cases} 1 \text{ if } \mathbf{X}-\mathbf{R} > \mathbf{0} \\ 0 \text{ if } \mathbf{X}-\mathbf{R} < \mathbf{0} \end{cases} \Bigg|_{\mathbf{R} \in \mathop{\times}_{V=1}^{T} \mathbb{R}} - \begin{cases} 1 \text{ if } \mathbf{P}-\mathbf{Y} > \mathbf{0} \\ 0 \text{ if } \mathbf{P}-\mathbf{Y} < \mathbf{0} \end{cases} \Bigg|_{\mathbf{P} \in \mathop{\times}_{V=1}^{I} \mathbb{R}}$$

$$= \prod_{V=1}^{T} \begin{cases} 1 \text{ if } X_V - R_V > 0 \\ 0 \text{ if } X_V - R_V < 0 \end{cases} \Bigg|_{(X_V , R_V) \in \mathbb{R}\times\mathbb{R}}$$

$$- \prod_{V=1}^{I} \begin{cases} 1 \text{ if } P_V - Y_V > 0 \\ 0 \text{ if } P_V - Y_V < 0 \end{cases} \Bigg|_{(P_V , Y_V) \in \mathbb{R}\times\mathbb{R}}$$

$$= \prod_{V=1}^{T} \text{hyp ( op log op (} X_V - R_V \text{ ) } / j \text{ , _ )}$$

$$- \prod_{V=1}^{I} \text{hyp ( op log op (} P_V - Y_V \text{ ) } / j \text{ , _ )}$$

$$= \prod_{V=1}^{T} \vartheta_{X_V - R_V ,1} - \prod_{V=1}^{I} \vartheta_{P_V - Y_V ,1} = \vartheta_{\mathbf{X}-\mathbf{R},1} - \vartheta_{\mathbf{P}-\mathbf{Y},1}$$

$$= \theta_{\mathbf{X}-\mathbf{R},1} \mid_{X_V - R_V \in \mathbb{R}/\{0\}} - \theta_{\mathbf{P}-\mathbf{Y},1} \mid_{P_V - Y_V \in \mathbb{R}/\{0\}} .$$

For instance: $\mathbf{X}$ is a set of inputs and $(\rho \mathbf{X})\, \varphi \mathbf{X} = \mathbf{Y}$ is a set of all those inputs plus all outputs which are labeled by invertible map $\rho$ after they are associated with them by process $\varphi$, namely, map $\varphi$ « leads » configuration $\mathbf{X}$ to point-location $\mathbf{Y}$ in phase-space $\mathop{\times}_{V=1}^{I} \mathbb{R}$ whose dimension $I = \dim \mathop{\times}_{V=1}^{I} \mathbb{R}$ and transformation $\rho$ of system $\varphi$ reversibly scales $\varphi \mathbf{X}$ there where this « naming convention » is really necessary for universal existence of $(\mathbf{R}, \mathbf{P})$.

If $\mathbf{R} < \mathbf{P} < \mathbf{X} = \mathbf{Y}$, then $\vartheta_{\mathbf{X}-\mathbf{R}\,|\,\mathbf{P}-\mathbf{Y}\,,\,1}$

$$= \prod_{V=1}^{T=I} \mathrm{hyp}\ \left( \log \frac{R_V - X_V}{Y_V - P_V}\ ,\ _ \right) .$$

Let $\mathbf{R} < \mathbf{X} = \mathbf{Y} < \mathbf{P}$. If so, then $\vartheta_{\mathbf{X}-\mathbf{R}\,|\,\mathbf{Y}-\mathbf{P}\,,\,1}$

$$= \prod_{V=1}^{T=I} \left[ \mathrm{hyp}\ \left( \log \frac{R_V - X_V}{P_V - Y_V}\ ,\ _ \right) = \mathrm{hyp}\ \left( \frac{1}{j} \cdot \int_{R_V}^{P_V} \frac{\mathrm{op}\ d\,y}{y - J}\ ,\ _ \right)\ \kappa \right]$$

(Sato 1959; Imai 1992). Thus we generalize switches and switching circuits (e.g., Klir 1972), but we can discuss discrete change and discrete events analytically and so strictly continuously.

## §3
## Procedures

[Much of this otherwise rather lengthy section and the next section, which derive demand for products from a category theory model of consumer preferences and behavior, has been edited out in this pre-print except for those parts relevant to section 5, which deals with managerial decisions and data entrepreneurs and investors are confronted with.]

Marginal quantities are combinations of replacements of one event in by another viewed from perspective of a definite consumer. Actors prefer certain events instead of other events and generally they wish to experience events in a particular order.

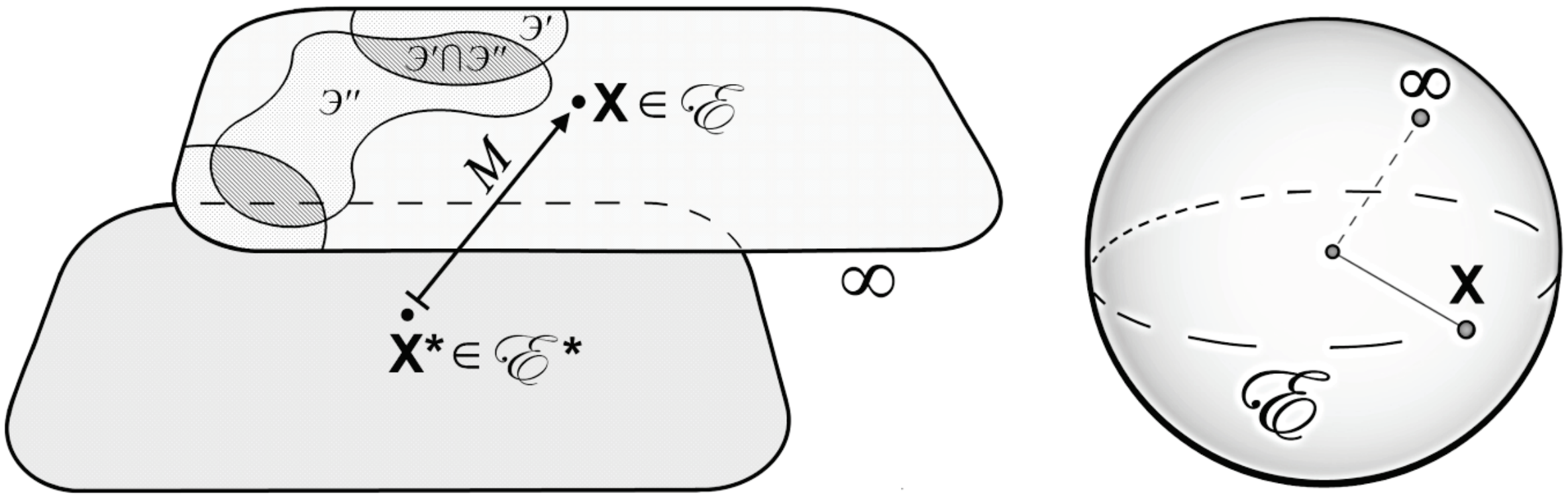

*Figure* 1. A marginal quantity $\bigcap_{Ц} Э^{ч}_{Ц} \in \times_{Ц} \mathscr{E}^{ч}$.

If various marginal quantities are uniquely but arbitrarily, consistently, and reversibly associated one-to-one with real numerical labels $\xi \in \mathbb{R}$, which are strictly hierarchically ordered,

then preferences of a consumer are generated by $\text{hyp}\,(F_+ , F_- )\,\xi$ there where

$$F_+ \;=\; F_- \;=\; \sum_{\mu} .\;\; \mathscr{R}\, J \cdot \zeta_\mu \cdot \vartheta_{J - R_\mu \,|\, J - R_{\mu + d\mu}\, ,\, 1} \;/\; j\,,$$

which is to say, in other words, that converging series that generate hierarchical preferences — over complex space restricted by a homomorphism $\mathscr{R}$ — can be decomposed into sums of projections of impulses which generate hierarchies of values in real space. This procedure can be physically interpreted: brains are functionally spectrum analyzers (Pribram 1991).

If $(N, n) \in \mathbb{C} \times (\mathbb{C}/N)$, then $\mathrm{hyp}\,(F_{+N}, F_{-N}) = \mathrm{hyp}\,(F_{+n}, F_{-n}) = F\xi$, or in other words, $(F_+, F_-)$ is not unique: infinitely many distinguishable pairs $(F_+, F_-)$ generate identical $F\xi$ — for instance: $\mathrm{hyp}\,(1, 0) = \mathrm{hyp}\,(1/2, -1/2) = \mathrm{hyp}\,(0, -1) = \ldots = 1$ — but $\mathscr{R}$ excludes those sets which are not order-preserving.

Neither $(F_A \cdot F_B)\xi$ nor $(F_A / F_B)\xi$ exist in general (Imai 1992; Graf 2010). This causes impossibility of inter-personal comparisons of preferences or values.

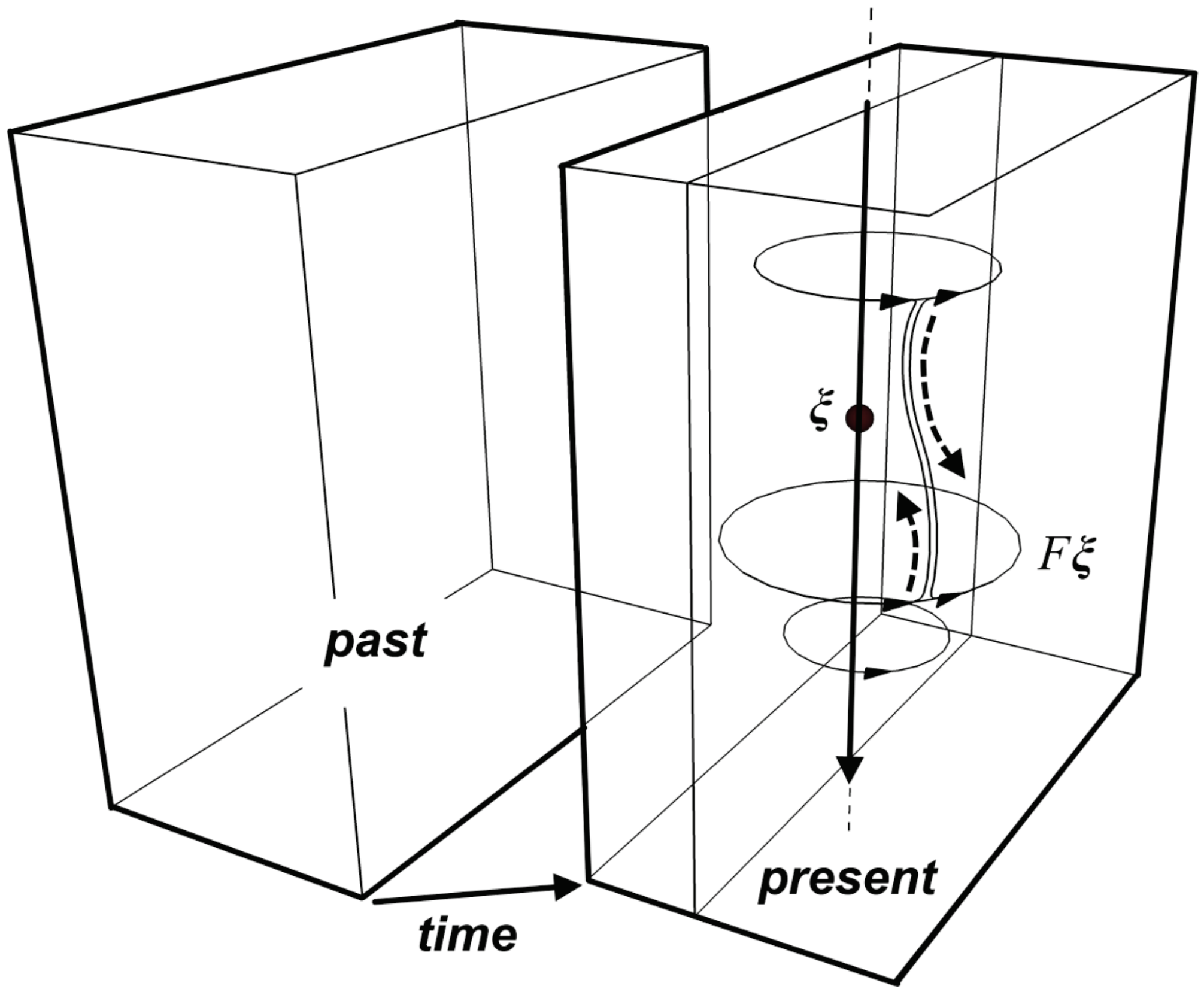

*Figure* 2. Following Imai's (1992) suggestion we can imagine $F\xi$ as complex velocity of series of various uniform distributions of sources and vortices which are spread around some path. Using Thomson's (1868) theorem we can visualize vortex layer $F\xi$ where complex flow of some infinitesimal element is identical everywhere along distinguishable circulations.

## §4

## Corollaries

Singularities $R_\mu$ are unique stimuli that don't evoke unique responses: conscious behavior is not caused by them but they determine reflex behavior. These experiences do not result in action; they result in reaction; *they do not cause choices*; but if so, then responses which are emotional, not conscious, not want-satisfying, reflexes, neither rational nor irrational, cannot contribute to economic behavior (Mises 1949).

Every person can perform unique motion in response to conscious desire to do so; but he or she can perform that motion by reflex, too, unconsciously, in order to remove one or other source of irritation, « emotionally » (Hebb 1949). Suppose that map $\Xi$ interrupts sensation of present stimulus $\xi$, stopping present action, replaces felt stimulus, and begins a different behavior by reflex. This response does not take place in markets: it does not determine choices that people make. Thus a complete system of behavior is:

$$\Xi = \sum_{\mu} . \ \rho_\mu \cdot \delta^{J}_{R_\mu} , \qquad F \, \Xi \, \xi = \zeta .$$

One popular criticism of economics is that it ignores such irrational behavior; but this criticism is quite invalid. We can ignore reflex behavior. It does not cause choices, and cannot therefore be invoked to explain economic behavior in markets (Mises 1949):

$$\frac{1}{j} \cdot \left[ \int_{R_\eta}^{R_{\eta + d\eta}} + \int_{\bigcirc(\eta, +)} + \int_{R_{\eta + d\eta}}^{R_\mu} + \int_{\bigcirc(\mu, -)} + \int_{R_\mu}^{R_{\mu + d\mu}} \right.$$

$$\left. = \int_{R_\eta}^{R_{\eta + d\eta}} + \oint + \int_{R_{\eta + d\eta}}^{R_\mu} + \oint + \int_{R_\mu}^{R_{\mu + d\mu}} \right] \frac{\text{op } d\,y}{y - J} = j \cdot Z + \frac{1}{j} \cdot \int_{R_\eta}^{R_\mu} \frac{\text{op } d\,y}{y - J} .$$

In other words, hyperfunction $F\,\xi$ sums around singularities $R_\eta$ or $R_\mu$ by way of arbitrary curves $\bigcirc(\eta, +)$ or $\bigcirc(\mu, -)$ that are closed, which is made possible by holomorphicity of $F_+ = F_- = G$ and implies that $j$ is added or subtracted $Z \in \mathbb{Z}$ times, depending on paths chosen during summation; <u>but these additions cancel</u> during hyp $(F_+ , F_-)\,\xi$ .

Neither $(F_A \cdot F_B)\,\xi$ nor $(F_A \,/\, F_B)\,\xi$ exist in general (Imai 1992; Graf 2010). This causes impossibility of inter-personal comparisons of preferences or values: **(1)** singular points of $F_A$ and $F_B$ do not everywhere coincide, and **(2)** graph $F_A$ generally consists of a space of ordered pairs that cannot one-to-one correspond to ordered pairs which comprise graph $F_B$.

Finally, many authors (e.g., Mas-Colell 1985) devote significant work to developing simplifying assumptions in order to model preferences. Why do they do this? There is one reason. Essentially they wish to be able to analyze preferences by taking partial derivates.

In reality, no special assumptions are necessary, and quantities and behavior can be treated as discrete where this is necessary, or useful, or simply the nature of real world data. We can arbitrarily differentiate $\mathsf{P} = \{\, \mathsf{P} \in \mathbb{C} \mid \Re\, \mathsf{P} > 0 \,\}$ or integrate $\mathsf{P} = \{\, \mathsf{P} \in \mathbb{C} \mid \Re\, \mathsf{P} < 0 \,\}$ complete system $F\,\Xi\,\xi = \zeta$ by way of analyticity of $F_+ = F_-$ :

$$\partial^{\mathsf{P}} F\,\Xi\,\xi \,/\, (\partial J)^{\mathsf{P}} = \text{hyp}\ [\,\partial^{\mathsf{P}} F_+\,\Xi \,/\, (\partial J)^{\mathsf{P}},\, \partial^{\mathsf{P}} F_-\,\Xi \,/\, (\partial J)^{\mathsf{P}}\,]\,\xi\,.$$

This leads to one further conclusion. Observe that predictions of some kind of market failure originating in models of consumer preferences or economic behavior of actors where such predictions are direct result of simplifying assumptions introduced originally in order to make possible calculation of partial derivatives in models tractable are most likely spurious, because really no simplifying assumptions are required in the first place and strange results produced by introducing simplifying assumptions are probably erroneous.

[Further parts of this section concerning change of labeling convention for marginal quantities, namely, monotonic re-scaling of labels, their lack of uniqueness, and other operations that prevent interpersonal comparison of preferences. These parts are omitted.]

## §5
## Fundamentally Uncertain Phenomena in Commerce?

In deterministic theories of production, profit opportunities exist to be found. They are not subjective but instead objective. Entrepreneurs obtain profit by way of arbitrage once they notice these profit opportunities, by directing attention to such circumstances. They can have but do not necessarily require ownership of factors of production or capital in form of money to purchase or rent such other capital-goods, because once they have discovered one or more profit opportunities, they can always obtain loans of capital, and pay these back with interest while themselves obtaining profit, which results from differences in prices at which otherwise identical marginal quantities are sold in one market by different sellers and purchases at such prices by different consumers due to ignorance. One buys such marginal quantities at the lowest price from one seller who is unaware of other prices these goods are sold elsewhere and sells at some price that is greater than that price but less than the greatest price realized in that market to all sellers are selling such marginal quantities at prices exceeding the said lower price.

This causes a single price to emerge in the market, removing error, namely, preventing scarce goods from being sold at prices below what people value them as, which prevents overconsumption and shortage of them relative to supplied quantities of other less valuable goods, while giving incentive to greater production of goods that are relatively more scarce by raising their supremum price in that market.

Profit arises in this perspective from noticing error and removing it by arbitrage. It essentially prevents waste of scarce factors of production by encouraging production of more valuable goods by raising their price relative less valuable goods, and lowering price of those other goods, discouraging their production, causing supplies of goods to correspond over time to anticipate changing preferences of consumers. <u>One question in this theory is: do profit opportunities always exist or can they be exhausted in principle and in practice</u>?

Profit in that theory arises from differences in prices of goods in times of greater plenty of such goods and prices of otherwise identical goods in times of greater scarcity of such goods, because they are in greater demand. Such differences arise due to purchasing at <u>certain</u> prices in the present and selling at <u>uncertain</u> prices in from perspective of present observers. If all future prices where known, then sellers of factors of production or consumption goods in the present could set price premiums in the present that exhaust price differences resulting profit.

This uncertainty must not be measurable, otherwise it would be treated as risk, calculated over time, and present prices of marginal quantities sold would be future prices of these marginal quantities discounted by such measurable uncertainty. Profit and loss would not result. (Knight 1921; Mises 1949)

Such uncertainty can originate exclusively in preferences of consumers. This is because all valuation is ultimately derived from valuations of consumers and the only non-quantitative and hence not measurable elements in markets are value hierarchies of consumers. Change in markets is driven by change of preferences of consumers or, more frequently, anticipation of such change, because investments and manufacturing processes begun now can only yield output later, and satisfy future preferences of consumers. These preferences must be anticipated by investors and entrepreneurs, which is ultimately why investment and entrepreneurial functions so often overlap in practice despite being theoretically separable.

Several challenges have existed in this theory. For instance, in practice, one can create a profit opportunity by offering something consumers find they like after they have been informed of its existence. This appears to be always possible.

Each action in practice is speculative (Mises 1949) and a source of uncertainty, and therefore, profit or loss (Hulsmann 2000). In practice this uncertainty appears to exist in even in absence of changes in preferences of consumers or shocks to the environment market participants find themselves in. Is there really a source of uncertainty which cannot entirely be removed in principle?

In production, one can also in practice always find ways to decrease cost of manufacturing something or giving service by discovering ways to reduce slack in execution of productive operations. Bottlenecks in production can be found whose existence was not suspected and these can be removed by additional investment of capital whose return may exceed cost of this investment. This is an additional source of uncertainty and thus a source of profit and loss, but these decisions are generally made by managers within a business rather than such decisions involving offering a different product to consumers.

Let us attempt to answer the first question before working on the second. Unlike ordinary functions, which are special cases of hyperfunctions, the pair $(F_+, F_-)$ <u>is not unique</u>. In fact, every hierarchy of values or set of preferences is generated in one of infinitely many ways. This is in addition to ordered numerical labels associated with marginal quantities being arbitrary.

Although consumer preferences generated in different ways are identical and choices they make in markets are identical, other behavior can depend on precise pair $(F_+, F_-)$ that generates their preferences. This is determined to large extent by homomorphism $\mathscr{R}$ . Ultimately, the minds of actors have one or other configuration, but although infinitely many different configurations result in actors making identical choices, their other behavior which is entirely rational can be different, and this would result in different market outcome.

For instance, one such difference of behavior of consumers is their attention to events, which is certainly determined by precise configuration of their central nervous system. Differences in attention are possibly due to different preferences but also two individuals with identical preferences can notice different events. This shall influence them to make different buying decisions, and in turn, such decisions result in differences in revenue from perspective of managers or entrepreneurs themselves, who must decide from such ambiguous data in preferences of consumers had actually changed. Their judgment is itself determined by what they notice. One way or another is becomes speculative regardless of how certain other parameters affecting their decisions are. This source of uncertainty, that identical preferences imply choices which are identical but not always identical behavior in general, always exists in the background of all productive activities.

This is often responsible for losses and possibility of costs exceeding revenuers always exists, therefore, even where preferences themselves of consumers and environmental shocks are somehow or by hypothesis perfectly anticipated by managers of firms which sell consumption goods. Prices of factors of production and capital goods are derived from value of goods directly purchased by consumers and so this source of uncertainty is transmitted along investment chains.

In fact, attention of managers and entrepreneurs themselves is determined by preferences and very likely by precise configuration $(F_+, F_-)$ that is selected from infinitely many possibilities, and so, behavior of agents of other firms are never quite certain. For instance, there is no certainty that two managers who have identical goals shall behave identically or make identical judgments, at least if only because they notice different events which then influence their actual behavior.

Usual models of preferences and behavior in economics derived from Fisher (1892) and Pareto (1906a) incorporate absence of uniqueness of labeling convention of marginal quantities, which implies that numerical labels that describe identical preferences are not unique and so cannot be compared, preventing interpersonal comparisons of value. This is also true of all hyperfunction models of preferences; but in those other models behavior nothing else remains that can generate uncertainty observed in reality, which is not merely lack of knowledge of true preferences of other actors and unobserved environmental shocks. In those other models uncertainty is then introduced extra-mathematically in order to explain observed profit and loss in situation where it would appear that all relevant parameters are known. This is not required in hyperfunction models of preferences, which supply reasons for such uncertainty and so have significantly more explanatory power in comparison to the already excellent analysis of Fisher (1892), which is however an equilibrium model, and does not explain profit and loss.

Two inferences can be made from this fact:

1) Investment risk of failure to receive returns exceeding costs during expansion of productive capability is likely underestimated.

2) Inventories of firms must be as flexible as possible in order to avoid sinking capital at opportunity cost for an extended period of time, because even if preferences of consumers do not change there is always the possibility that a steady stream of purchases shall be temporarily be interrupted for reasons outside control of the manager and outside the predictions of the entrepreneurs. Rigid inventories are a source of loss in this context or at least foregone profit. This also suggests that profit opportunities exist where they are not expected and that some profit opportunities which have not been noticed always exist despite observation of other profit opportunities.

3) Cash flow must be sufficient in order to for a firm to be able to take advantage of a profit opportunity once this is found; otherwise it suffers an opportunity cost. This is necessary because some profit opportunities are entirely unexpected fundamentally and there are always such opportunities. Readiness in order to earn profit in these cases is required and there is always a risk of lacking required cash flow. In these cases, opportunities can be taken advantage of by way of borrowing but this is at cost of interest, which limits the set of profit opportunities that can be acted upon. Such lower limits can be calculated by way of a discount model having form:

$$\vartheta_{\mathbf{X}-\mathbf{R}\,|\,\mathbf{Y}-\mathbf{P}\,,\,1} \cdot (1 - r \cdot \delta_{\mathbf{N}}^{\mathbf{A}\,,\,0} \cdot \delta_{\mathbf{M}}^{\mathbf{B}\,,\,0} \cdot \ldots\,).$$

These considerations relate to the second question posed earlier which concerned reorganization of production to lower costs. One of the fundamental questions facing managers of investment chains is the issue of making decisions based on data which is partly erroneous or inaccurate to some degree, which fact is complicated by presence of usually many different sources of lack of precision or inaccuracy, which creates the issue of procedure of quantitative calculation of using such data (Morgenstern 1950). Such data may consist partly of estimates of results of an activity which may take place in the future instead of actual outputs themselves.

Most data whose accuracy is limited is represented by an interval. For instance, a producer does not know precisely true revenue in the next month with any certainty, but he or she is almost entirely certain that it shall be within such and such range. Let us suppose as is often the case that he or she makes a present judgments not on the precise revenue but rather in what interval that revenue is anticipated to be. If so, let us desire to measure, during a calculation the contribution to revenue of varying some other factor of production whose quantity is certainly know. Let us suppose again, as is usually the case, that several decisions made in the past were determined by interval considerations or estimates, too, and not precise quantities. Use of conventional intervals, discrete jumps and switches is possible (e.g., Klir 1972) but all common procedures would not allow calculation of partial derivatives, which implies that marginal contributions of various factors or events or decisions cannot be calculated on a general technical level there where mixes of discrete phenomena and data of imperfect accuracy exist.

In fact this is actually not a problem: $\theta$, $\vartheta$, and $\delta$ hyperfunctions constructed earlier in this paper and used throughout it admit inputs which are intervals and behave discretely where necessary without loss of analyticity. Thus a mixed set of data such that which managers and entrepreneurs observe in practice is always tractable to calculation of contributions of various events, including unique events and estimates, to revenue, even where this is non-linear.

In this way anticipated returns to various activities, which are generally speculative and known or anticipated with limited degree of precision and certainty, can be manipulated as quantitative data and compared with other such decisions. Estimation of risk would involve models comprising mostly of maps of type $\delta_{\mathbf{R}}^{\mathbf{X},0}$ as inputs to technical models of process, whose change in response to certain rare events can thus be estimated in turn and discounted. Results of numerous managerial and entrepreneurial decisions thus become open to analysis.

In general, a manufacturing process or business plan or estimate of activity can be treated as process where data table $\mathbf{X}_0$ consists of zeroes and greater than zero coefficients, and when sufficiently many coefficients that become greater than null, e.g., $\varphi \mathbf{X} > \mathbf{R} > \mathbf{X}_0$, a certain amount is added to revenue or subtracted from revenue and the equations which determine this process permanently change (Shapiro 2013). This amount is discounted by continuous interest determined by market interest rate over the interval of time beginning with start of work on that process and interval estimate of risk or probability of this event taking place (Hayek 1941; Mises 1949).

Although revenue uncertainty always exists, all planning of firm activities can be quantitatively specified by incorporation of estimates of future values of active assets and results of past discrete actions into quantitative models of firm revenue that already incorporate other numerical variable quantities. This may be useful in order to limit the opportunity cost associated with action under uncertainty through projections of revenue which make use greatest possible use of all available data while permitting use of all tools of rigorous analysis.

§6

## Examples

§7

# Part 3. Several Corollaries of Procedures of Management of Manufacturing Processes. Are Capital-Based View of Production & Knowledge-Based View of Production Alternate Descriptions?

**Abstract.** If marginal-productivity gains by way of increase of extent of division of labor in knowledge and work are observed from perspective of entrepreneurs or managers, namely, endo-physically — from perspective of parts of those systems themselves whose behavior is studied — which is how all manufacturing processes are observed in actual practice by those who participate in them, then Capital-Based View of Production (CBVP) and Knowledge-Based View of Production (KBVP) are found to be alternate descriptions of identical underlying manufacturing processes. CBVP is not reducible to KBVP and KBVP is not reducible to CBVP. Use of knowledge in manufacturing processes is discussed.

## §1
## Introduction

Several models can be alternate descriptions of identical underlying natural processes. This results from existence of multiple fundamentally different ways of interacting with these systems, which is true if they are complex systems, and indeed their complexity or extent of hierarchical organization is measured, for instance, by logarithm of number of such alternate descriptions that are not reducible one to other, so that processes whose behavior is exhaustively mapped by one procedure of interacting with it — are simple categories — have null complexity. (Rosen 1969; 1975; 1985)

Valuing something implies anticipation that it satisfies one or several wants by way of its consumption or contributes to production of something that does so (Menger 1871). Whatever cannot satisfy wants and does not contribute to happiness cannot possibly be wealth; that consists of things which actually do satisfy wants (North 1691; Bonnot de Condillac 1776; Say 1803; Craig 1821). If all marginal quantities or goods have unique numerical labels, then from perspective of person $Ч$ « order » $\mathcal{O} \in \mathbb{R}$ of « event » $\xi \in \mathbb{R}$ is countable measure of its distance $Д(Ж, \xi) = Д(\xi, Ж) = \log \mathcal{O}$ from ultimate cause of value of goods, which is anticipated or actual want-satisfaction felt when goods are manipulated by that observer, namely, « consumption » $Ж$, where $\mathcal{O} = 1$, because $Д(Ж, Ж) = 0 = \log 1$ (Menger 1871).

A manufacturing process begun at one moment — from a perspective — can possibly yield completed product only later from that perspective, and this product is proportional to degree of extent of division of labor or length of investment chains involved — or, in other words, « roundaboutness » of that manufacturing process — whose union is that process of production, and greater division of labor results in greater marginal productivity — more output per input — but later, too, and must be discounted while being done; such increase is limited by extent of market demand for products of these investment chains (Longfield 1834; Mises 1912; Hayek 1941; Shenoy 2007). This increase of extent of division of labor is done by way of change or re-organization of processes of manufacturing that requires additional capital and time, for instance, beginning higher-order tasks while reducing number of lower-order tasks, or possibly by greatly increasing number of lower-order tasks while slightly reducing number of higher-order tasks, or beginning higher-order tasks of while also increasing number of lower-order tasks; additional division of labor of knowledge or work requires additional time to organize and additional capital to execute, and ultimately produces more economic goods than can possibly be made without it, discounting risk of failure to manufacture, but these are necessarily made available later, and this is lengthening of investment chains, at least organizationally, and decrease of market rate of interest (Huerta de Soto 1998), which is ratio of part of instantaneous income that is consumed to remainder of instantaneous income that is invested, or if that is measured by money-prices, ratio of consumption to capital plus risk (Mises 1912; 1949).

Thus several observers (e.g., Fisher 1930) have erroneously inferred that investment opportunities — marginal productivity — determines personal rates of interest together with time-preferences of consumers, instead of them being exclusively determined by time-preferences of consumers (e.g., Mises 1912; 1949; Fetter 1914a,b), who at every successive moment spend part of their income on immediate future or present consumption, including common medium of exchange or its substitutes held — in order to easily or quickly purchase economic goods later despite present uncertainty of what they later want to buy — and save and invest remainder, which determines later future income and so later consumption from their perspective (Craig 1821). Why? This is so because marginal productivity is already paid for by rent or purchase of factors of production in the market, whose value is determined by their contribution to value of the future product assembled from them discounted by time lapse.

Producers must anticipate what methods of production shall actually satisfy consumer preferences, which they do not know and must anticipate too because they are in the future, when it is already too late to do so if appropriate production is not begun in the present; all productive behavior is thus doubly anticipatory or speculative (Mises 1949).

Equivalence relation of cause and effect necessarily implies that positive work is result exclusively of positive energy and change of state: it takes place over time, and so investment and production begun in the present can only yield products in the future, and products which are available later but not now possibly satisfy exclusively future wants of investors or savers, not their present wants. Production begun in concrete moment $H$ possibly satisfies exclusively future preferences in future moment $K$. Thus goods that don't exist in some moment of time — for instance because they were not produced earlier from perspective of consumers — cannot be consumed then and there.

Market prices allocate information — by way of indexing not measuring — not possibly available to any single entity regarding production decisions (Gossen 1854; Hayek 1945); but entrepreneurial judgment, although guided by such prices, is anticipatory or speculative (Knight 1821; Mises 1949). This fact causes profit or loss by way of actors purchasing at certain prices earlier and then necessarily selling at fundamentally uncertain prices later (Cantillon 1755; Turgot 1769; Knight 1921; Mises 1949; 1952).

Observe that degree of progress of welfare of society is directly proportional to degree of progress of human knowledge of causal connections between concrete things and welfare, namely, satisfaction of human wants, because in order to satisfy human wants, one must correctly anticipate or know what things, by way of their consumption or use, can actually do so or contribute to such activity by transforming into parts of such directly useful things (Menger 1871). Accumulation of recipes of identity of combinations or separations of things followed by manufacturing output by so transforming input does not exhaust possible knowledge. Perception of greater usefulness — possibility of satisfying additional wants of somebody or contributing to such action — of things already possessed are gains of knowledge too.

In absence of new technological knowledge some processes or things which are not used in production always exist already but scarcity of capital and other alternative ways of using income causes opportunities that must be foregone in order to use them, namely, their additional cost, to exceed additional revenue they would generate. Making, purchasing, or renting such additional higher-order goods in order to manufacture additional lower-order goods would lengthen structure of production so greatly that productivity gained by using them, which is ultimately indexed by ratio of additional revenue they generate and their additional cost, is less than rate of interest and using them would result in losses instead of possibility of profits. Let supply of capital increase and rate of interest decrease and entrepreneurs would easily find and begin using new techniques or new tools that result in greater output per supply of input. This is true if technological knowledge is stationary, and so, necessarily, it remains true while additional technological knowledge is learned and accumulated. (Bohm-Bawerk 1909)

Observe that knowledge is true knowledge: one cannot know a given proposition, which relates causes and effects (Leibniz 1695), that isn't true, but one can believe a false proposition or a true proposition. If so, then knowledge, in this sense, is a resource in manufacturing processes, part of set of capital goods, because it requires time to create, by saving and investment, and yields better ability to judge marginal productivity or value of *N*-th order goods. Knowledge is a resource that can be had more or less of, a kind of capital good, because value can be correctly or erroneously anticipated without true knowledge of causality by anticipating causality by other ways (e.g. guessing or speculative judgment), since all human action is speculative and uncertain: knowledge reduces uncertainty in results of production and leads to greater frequency of choices resulting in action that actually can satisfy wants they were anticipated to satisfy — « rational » decisions — given all set of decisions acted upon (Mises 1944; 1949; Hulsmann 2000).

While ends are subjective, and cannot be rational or irrational; but means are not such; they can be rational or irrational, which depends on whether do, in fact, satisfy wants they were intended to satisfy, whether they success or failure of action (Mises 1944; 1949; Hulsmann 2000).

To satisfy wants, one must anticipate or perceive causality between what satisfies wants by way of unproductive consumption or contributes to making something that satisfies wants by way of productive consumption, get it, and consume it appropriately.

Actually, perception can be erroneous, and anticipation can diverge from reality, and sometimes, one can act rationally or choose that action which actually can satisfies those wants not for any valid reasons by for erroneous reasons, merely be coincidence of other circumstances outside one's control (Hulsmann 2000); that rationality or irrationality is not purely environment determined but always contained in every action partly, and rational action is possible in every environment; whether it actually happens and how often is another question. How? In same way that one can believe a true proposition and that it is a true proposition without knowing that it is a true proposition, or concisely, « without knowing it », because one's reasons for believing it, unknown to one, are not actually true, merely possibly true or sometimes necessarily impossible even (Gettier 1963).

In other words: one can know a true proposition without sufficient reason why it is true or infer by false inference a true proposition from a false one. Therefore one does not always need to be justified, or know true reasons for something, in order to believe that something and for that something to be true. See also Hulsmann (2000) in this connection: one can act rationally, satisfy wants intended to satisfy, without knowing causality or true reason why such action satisfies want, or erroneously perceive causality but for other reasons outside of one's control still have one's wants satisfied. Of course: *frequency* of having wants satisfied this way is much less than such *frequency* there where true causality and true reasons why action is rational is known and thus action is likely — except for other circumstances outside one's control — going to satisfy one's wants (Bohm-Bawerk 1889).

How is human knowledge related to production of that which satisfies human desire? Some writers have said that knowledge is power itself. « Nature is not used except by obedience. People can do nothing except move and remove natural bodies. Nature performs all else; human knowledge and human power coincide. Ignorance of cause hinders production of its effect. » (Bacon 1620)

This is not everywhere true; and why? Frequently true knowledge of cause and effect in nature is not sought exclusively for power but because gain of it is pleasant to people who seek it (Mises 1949).

This knowledge is otherwise really a capital good itself — a higher-order good — but it is one of several higher-order goods and so where other capital goods complementary to it in production of lower-order goods are lacking in supply greater quantity of things which satisfy human wants or can contribute to their satisfaction — greater « income », « wealth », or supply of first-order goods (North 1691; Craig 1821; Menger 1871) — cannot be made (Mises 1949; 1952). All true knowledge of cause and effect without opportunity to use it to contribute to manufacture of things which satisfy preferences of somebody cannot itself satisfy his or her preferences unless he or she prefers immediate possession of more true knowledge than less knowledge of truth; but only philosophers — lovers of truth — have such preferences; most people value truth only where it is anticipated to ultimately used to contribute to production of first-order goods, namely, consumption goods (Mises 1949).

This especially takes place where causes are global but effects are local and human action is always local and cannot change such causes whether we know or not that they result in such and such effects (Thom 1983), e.g., a man during a flood drowns both if he knows what caused the flood and if he does not know this, and like we can fully know relation of cause and effect but fail to use this knowledge to satisfy our wants so we can believe false causation or lack knowledge of it altogether and by chance or trial and error and repetition satisfy our wants where this is possible, e.g., medicine; global causality which we cannot change is <<evident in social interaction>> (Thom 1979; 1983). Both opportunity and action are required to satisfy wants: one without other cannot do so (Davenport 1918). This is part of law of correspondence of state of interior of a system to state of its exterior which impinges on it and would destroy it if state of its interior did not compensate for state of its exterior (Spencer 1864).

Plato said that philosophers are those persons who love truth, which is to say, in other words, that they do not necessarily already know it. Why did he say that? Men and women cannot possibly already possess all true knowledge — know every relation of cause and effect — that can possibly be possessed because such knowledge is not finite and so all they can do is seek more of it if they want to have it. Hutton (1794) wrote that scientists are those persons who do what is required in order to learn truth and his friend Smith (1776) had therefore earlier written that wrote that philosophers are those persons who observe everything and do nothing; of course, those who do nothing to satisfy wants of other people are not given or paid anything in exchange for this non-service by other people, which is one truth that scientists quickly learn by experience.

Technical knowledge is closely related to production since it is used by organized manufacturing processes and it consists of patterns. These are cross-sections of fiber-spaces (Rosen 1981).

A question is here posed: what types of fiber-spaces are involved in production of goods? Attempt shall be made to answer this question in later sections of this paper. Here we contrast two methodologies of answering this question.

One is called the Knowledge-Based View of Production (KBVP), in which analysis knowledge is a capital good, treated a resource possessed by owners of manufacturing processes in some quantity, and the Capital-Based View of Production (CBVP), in which knowledge is perceived not as resource itself but as a dimension of all capital goods, because knowledge of cause and effect is involved in valuing all capital goods and so cannot be a capital good itself (Baetjer & Lewin 2011).

In both methodologies changes in knowledge evoke corresponding additions to capital involved in manufacturing processes, but it would certainly appear that from perspective of accounting and management these are highly opposed views. A manager who follows the KBVP shall attempt to purchase knowledge and shall try to measure its quantity and part of revenue which results from sale of product of a manufacturing process shall be imputed as marginal contribution to value of that product created by that knowledge and its source will be paid rent accordingly or, at least, this shall be treated as a cost of production. This is not so in the CBVP. There a gain of knowledge shall be considered as a gain of capital but this gain is achieved by learning done by entrepreneurs and its cost is opportunity cost of time devoted to such learning, not a market price of a factor of production that is bought and sold or whose lowest price in the market is determined by it marginal productivity. Instead it adjusts valuation of all other goods of greater than first-order, i.e., it increases or decreases estimation of capital possessed by a business from perspective of whether these higher-order goods contribute to value of products, which itself is, too, estimated according to such knowledge.

I shall argue that these views are complementary, that both costs exist, and that knowledge is used in both senses in production, by way of modeling of production processes.

## §2

## Notation

[ Refer to a basically identical section in my paper: « Leibniz's Law of Continuous Change and Analysis of Fundamentally Uncertain Phenomena in Commerce. » There is one difference here. ]

Number of different space-time events all labeled, marked, or named $L$ — elements in set $L$ — is countable measure $\sum_P \sum_R \delta_R^{P,1} = \sum_P P^0 = \langle L \rangle \in \mathbb{N} \cup \{0\}$ (Jevons 1870; 1874; 1880), or in other words, events whose « quantity » by hypothesis is neither one nor zero are necessarily distinguishable in quality by their behavior or by their other marks, too, because events which do not so differ are one thing contrary to hypothesis that they are several; but that is self-contradiction, impossible, and so they cannot possibly differ numerically — they cannot be labeled by non-unit « quantities » — without differing in « quality » (Leibniz 1717). Selected part $P \in \{ \underbrace{\eta, \ldots, \mu}_{\langle L \rangle} \} = L$. Ordered pair $(R, P) \in L \times L$.

Permutation $\bigtimes_P = \underbrace{J \times \cdots \times J}_{\langle L \rangle} = \bigtimes_P . \; J + 0 \cdot P$ . Dimension $\langle L \rangle = \dim \bigtimes_P$ .

## §3

## Procedures.

Production involves overcoming over time natural resistance to change of form of things; it is neither creation nor destruction of things but change of their form or position by way of interaction with them; consumption corresponds to opposite change of their form (Say 1803; Gossen 1854). Thus manufacture of means of satisfying wants is « productive » consumption but satisfaction of wants itself — ultimate end of purposeful activity whose possibility is sole original source of value of such means — is « unproductive » consumption (Longfield 1834). Many types of « unproductive » consumption necessarily take place over time, not instantaneously, too, like « productive » consumption (Hayek 1941).

Work $\mathbf{W}$ in order to transform inputs into outputs must equal such resistance $\mathbf{R}$. This is necessarily greater than zero assuming positive production is done from a producer's perspective during time interval $\mathbf{B}$. Oriented « bridges » or « jumps » across energy « gaps » or potential changes of state possible but not actually taking place for lack additional energy — or directed « tunnels » passing « through » elsewhere infinitely resisting « barriers », allowing materials to flow along « value » gradients, and determining frequency of things moving from volumes of lesser marginal utility to volumes of greater marginal utility — are manufacturing processes. Width of these surfaces, multiplied by their infinitesimal height, limits or restricts frequency $\phi$ of event corresponding to real numerical label $\xi$.

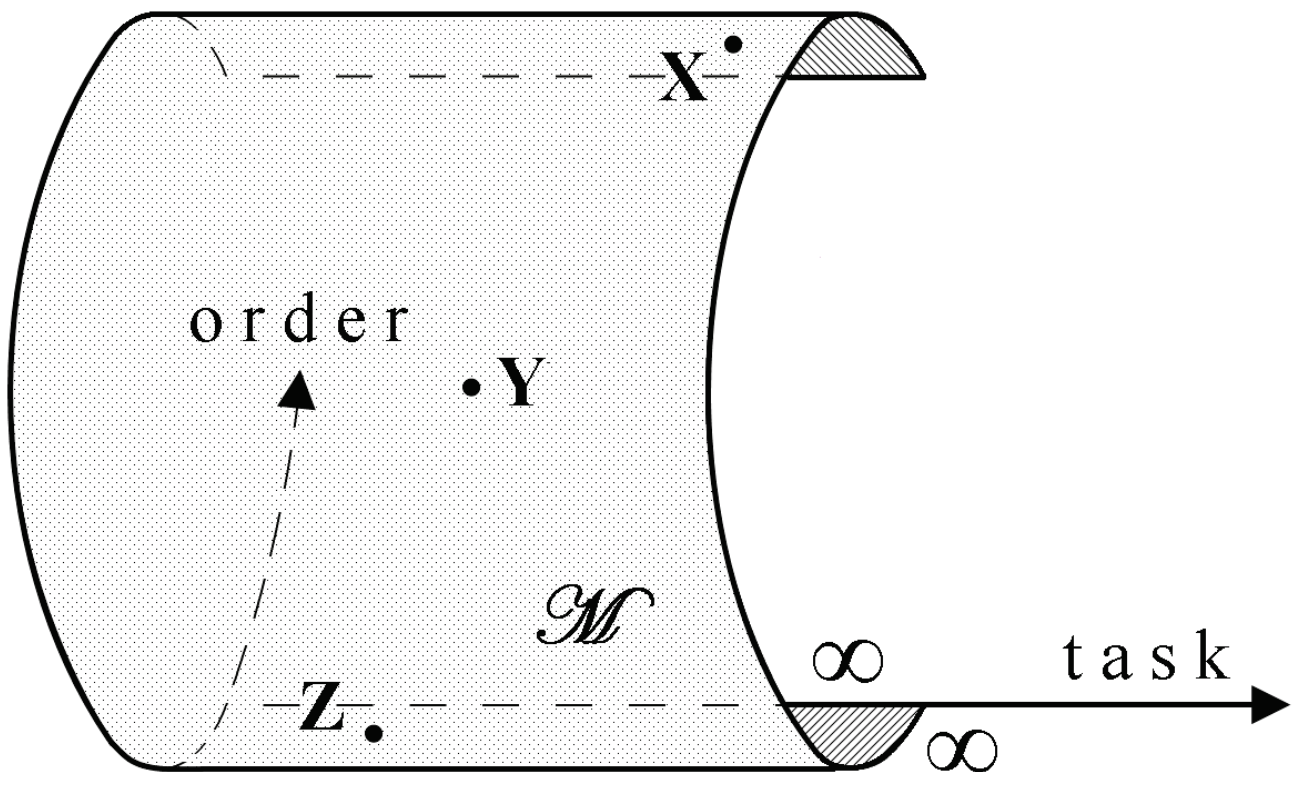

*Figure* 1. (Order, Task)-Manifold $\mathscr{M} = \mathbb{R} \times \mathbb{R}$.

Thus infinitesimally thick sheet $IK$ moves across energy gap $i \cdot (\omega + \omega)\mathbf{X} = \mathbf{R} = \mathbf{W}$ relative to origin $\mathbf{0}$. Set of every physical time moment $K$ from perspective of person $Ч$ is orthogonal to space that consists of every ordered triple $(\mathscr{O}, \xi, Ч)$.

Value of higher-order is derived from those wants which they contribute to satisfying, e.g., their present prices are their time-discounted marginal contributions to future prices of lower-order goods which are ultimately later assembled from them by way of their productive consumption in order to eventually make first-order goods which satisfy wants by way of being unproductively consumed by their ultimate consumers (Longfield 1834; Menger 1871; Mises 1949). Every manufacturing process consists of an ordered combination of « tasks », one or several consumptive operations that are productive (Longfield 1834), higher-order goods (Menger 1871), « active-assets » or « capital-goods », sum of whose present money-prices, namely, « capital », is compared with total revenue which they generate; or in other words, a positive difference of money-revenue and money-cost is a « profit », and a « loss » — which limits productive consumption of higher-order goods in one manufacturing process and redirects them into hands of some other producer is a negative difference of money-revenue and money-cost (Menger 1888; Mises 1949; 1952). A manufacturing process thus consists of set $\{\xi \mid Д(Ж, \xi) > 1\} = \{\xi|_{\mathscr{O} > 1}\}$.

One possibly redundant path $\mathscr{P}$ over positive time, interval $\mathbf{B}$, from perspective $Ф$, along directed edges of graph of ensemble $\mathscr{E} = \{\xi|_{\mathscr{O} > 1}\}$ of productive tasks $\{\xi|_{\mathscr{O} > 1}\}$ — operators which change form of inputs — is a manufacturing process executed by firm $Ф$. Such operators are usually non-commuting but this isn't necessary everywhere in space-time. This ensemble is completely partitioned by intersecting it with disjoint sets $\mathscr{A}'$, $\mathscr{A}''$, …, whose elements are possible investments of additional capital that are not co-possible, « alternatives », that exhaustively cover it: several such coverings usually exist. Procedures of discovering limits of manufacturing processes — their « bottlenecks » — and choosing actual investments from sets of alternatives so that investment of additional capital results in greatest possible returns are necessarily methods of transforming capital goods into consumption goods with least forgone opportunity, namely, with least « cost ».

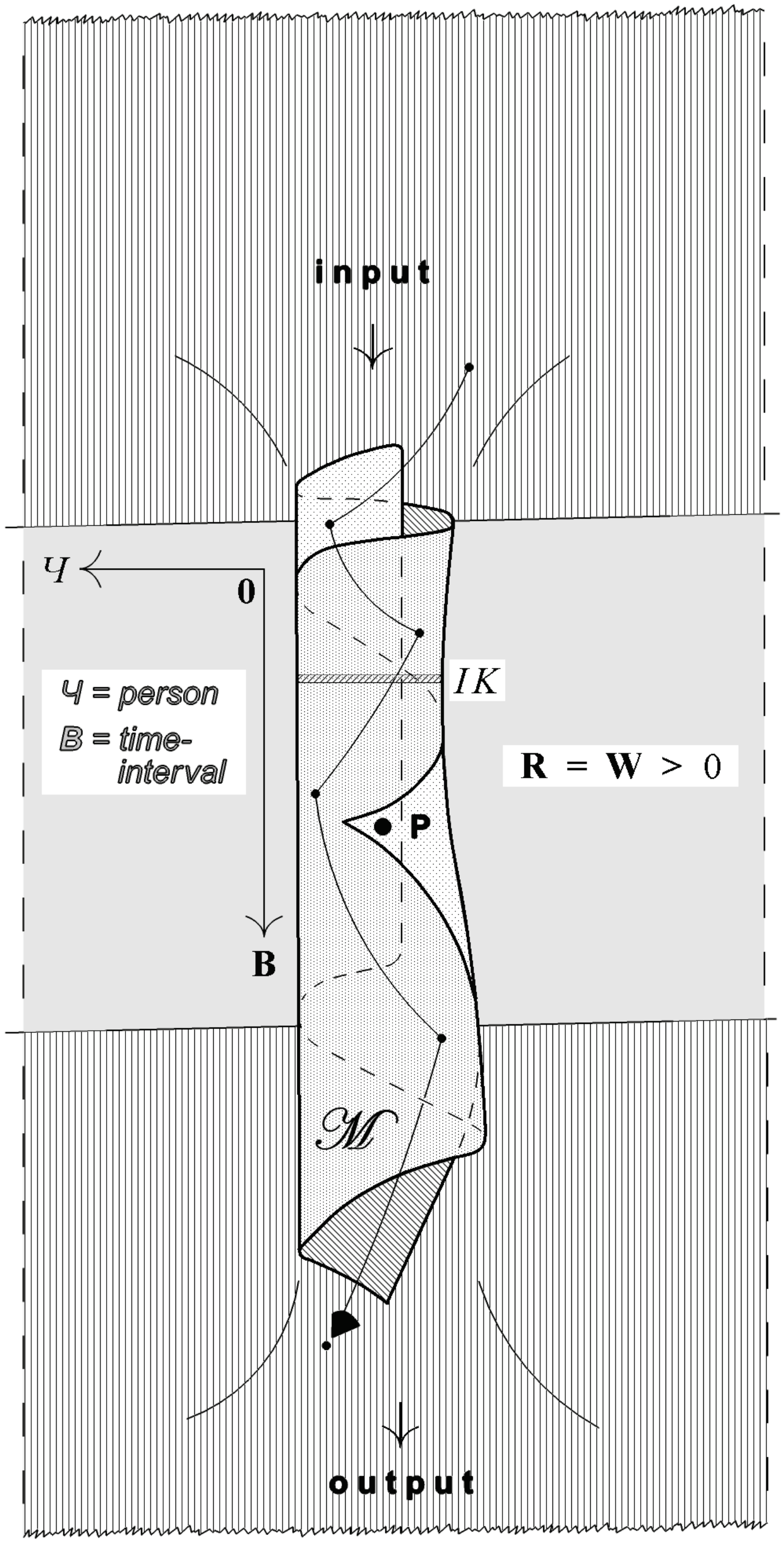

*Figure* 2. Folding of real manifold $\mathscr{M}$ according to form metric $C$ constructs structure of manufacturing process $\mathscr{P}$. Here several different tasks $\mathbf{X}$, $\mathbf{Y}$, $\mathbf{Z}$, which are completed during interval $\mathbf{B}$, are performed by one person at point $\mathbf{P}$. Why? $C(\mathbf{X},\mathbf{Y}) = C(\mathbf{Y},\mathbf{Z}) = C(\mathbf{Z},\mathbf{X}) = 0$ in phase-space $\mathbb{C} \times \mathbb{R} \times \mathbb{R}$.

Observe that all « tasks » $\xi$ have a « specificity » $\langle \exp Д Ж \xi \rangle$, or in other words, more specific goods are productively consumed in a great number of distinguishable stages of production than less specific goods (Hayek 1931). Why? $\exp Д Ж \xi = \exp Д(Ж, \xi) = \mathscr{O}$. Boundary $\partial \mathbf{B} = [H, K]$. Period time-preference is $\mathscr{I}^{K}_{H}{}^{*} = IK - IH > 0$. Let $\tau \in \mathbf{B}$. Instantaneous ratio of income unproductively consumed relative to income productively consumed is momentary time-preference $\mathscr{I}^{K}_{H} = F_{\mathbf{B}}\{\xi|_{\mathscr{O}=1}\} / F_{\mathbf{B}}\{\xi|_{\mathscr{O}>1}\} > 0$, which implies monotonic operator $\mathscr{D}^{K}_{H} = J \cdot \mathrm{op} \exp(\mathscr{I}^{K}_{H} \cdot (K - H)) = \mathscr{D}_{\mathbf{B}}$, namely, « discount », where

$$\log(1 + \mathscr{I}^{*K}_{H}) = \log \lim_{\tau \to \infty} (1 + \mathscr{I}_{H}^{K} / \tau)^{\tau} = \log \exp \mathscr{I}_{H}^{K} = \exp \log \mathscr{I}_{H}^{K} = \mathscr{I}_{H}^{K}$$

(Bohm-Bawerk 1889; Mises 1912; 1949; Hayek 1941; Huerta de Soto 1998).

Suppose that progressive increase of this ratio takes place for some reason, e.g., a change of time-preferences. If $\{ [\exp Д Ж \xi|_{V|\mathscr{O}>1}] - [\exp Д Ж \xi|_{W|\mathscr{O}>1}] > 0 \mid \langle \exp Д Ж \xi|_{V|\mathscr{O}>1} \rangle - \langle \exp Д Ж \xi|_{W|\mathscr{O}>1} \rangle > 0 \} \neq \varnothing$, then less shortening of investment chains or structure of production by way of decrease of capital stock or increase of time-preference is sufficient to cause people to entirely cease using type-$V$ goods in production, which causes them to lose value completely, than for type-$W$ goods, and during progressive shortening of structure of production, people abandon type-$V$ goods earlier than type-$W$ goods (Hayek 1931; Mises 1949).

If quantity of higher-order goods earlier used in production of lower-order goods is greater, then quantity of product is greater, too, because their combination is ultimately physical identical to such product; but this is discounted by risk of failure of manufacture by way of inevitable or fundamentally unforeseeable causes, and by time waiting for product. Thus greater future income results from greater part of present income productively consumed at cost of income possibly but not actually unproductively consumed. Instantaneous rate of interest $I$ at any moment for any arbitrary period is greater than zero, otherwise people would not consume, by putting off consumption forever, which is contrary to hypothesis of them having preferences, i.e., that people strive to satisfy wants, which can be done solely by consumption.

All change takes place over time and so production, too, necessarily takes place over time (Menger 1871). Motion of machinery that automatically displaces something is ultimately controlled by a person: every task is actually ultimately performed by a person (Mises 1949). Flat surface $\mathscr{M}$ is folded according to definite metric $C^{\Phi}$:

$$( \mathscr{O} , \xi ) \overset{C^{\Phi}}{\mapsto} ( Ч , K ) \in \mathbb{R} \times \mathbb{R} .$$

Let $\mathbf{B} = [ B^H , B^K ]$. Several reasons for greater productivity of division of labor exist: Smith (1776) effect (learned increased dexterity by repeated and specialized doing of particular behavior resulting in absolute advantage and natural absolute advantage), Bastiat (1851) effect (minimum cooperation required to overcome natural barriers causing absolute advantage), and Torrens (1815) effect (comparative advantage), and these all cause division of labor in society. To not do so and engage in isolated labor would present a significant opportunity cost in work. Such division of labor takes place in knowledge and work (Smith 1776).

All change takes place over time and so production, too, necessarily takes place over time (Menger 1871). This is worth repeating. There is further the fact that in order to organize additional division of labor requires additional time (Menger 1871). This is ultimately what is responsible for the often disputed technical superiority of present capital goods versus future capital goods, namely, investment begun earlier can be used in those investment chains where division of labor is more extensive all other things being the same (Bohm-Bawerk 1889). Thus division of labor, as is well-known, is limited by two factors, which are interest-rate determined by time-preference, since this continuously discounts greater product of more roundabout production, which limits its length (Hayek 1841), and extent of market demand for greater product, since a greater product is valuable only if there is demand for it and it can be sold at prices exceeding cost of making it (Smith 1776; Longfield 1834).

Let $(\xi, \sigma) \in \mathbb{R} \times \mathbb{R}$. Map $\underset{N}{\times} \mathbb{R} \xrightarrow{\omega \mathbf{X}} \mathbb{R}$. If $\xi - i \cdot \omega \mathbf{X} = \varsigma \in \mathscr{C}^{C} \subset \mathbb{C}$, then $(\varsigma + i \cdot \sigma, \varsigma - i \cdot \sigma) = (\kappa, \bar{\kappa}) \in \mathbb{C} \times \mathbb{C}$. Let $\mathsf{P}_{\mu + d \text{ abs } \mu} = \mathsf{R}_{\mu \in \mathbb{R}} \in \mathbb{R}$.

Resistance to change of form is a complex-volume with cross-section $\omega \mathbf{X} - \text{op } \omega \mathbf{X}$.

$$\text{Let } \quad \underset{H}{\overset{K}{\psi}}{}^{\mathscr{O}}_{\xi} = \text{hyp}\,(\psi^{\mathscr{O}}_{+}, \psi^{\mathscr{O}}_{-})\,\kappa = \sum_{\mu} \begin{cases} \psi_{\mu} & \text{if } \mathsf{P}_{\mu} < \xi \cap \xi < \mathsf{R}_{\mu} \\ 0 & \text{if } \mathsf{P}_{\mu} > \xi \cup \xi > \mathsf{R}_{\mu} \end{cases} \mid [\exp ДЖ\, \xi = \mathscr{O}] \quad .$$

$$\text{If so, then } \psi_{+} = \psi_{-} = \sum_{K} \sum_{\mu} \left[ \frac{\psi_{\mu, K}}{j} \cdot \log \frac{\mathsf{P}_{\mu, K} - J}{\mathsf{R}_{\mu, K} - J} \right].$$

An unfolding distribution results over time. Such construction of « hyperfunctional » but not « functional » maps (Kothe 1952; Sato 1958; 1959; 1960), is done in order to realize fundamentally discrete nature of productive choices and manufacturing decisions, namely, « investment opportunities », set of actions that increase future income discounted by cost of waiting and thus excludes pure loans (Fisher 1930).

Frequency $\underset{H}{\overset{K}{\phi}}{}^{\mathscr{O}, \Phi}_{\xi}$ is thus determined by folding map $C^{\Phi}$ of discrete choices-decisions made during manufacturing by consuming-producing entity or firm $\Phi$ . Fisher (1892) suggested that production, which is fundamentally discrete but manifests in flows, must be analyzed by considering frequency of various kinds of events, which is a procedure invented by Waterston (1831), in order to determine mean behavior of some system of particles by determining frequency of various types of collisions. By itself this type of analysis is not sufficient and requires supplementation.

## §4
## Analysis

[ Refer to related sections of: « Leibniz's Law of Continuous Change and Analysis of Fundamentally Uncertain Phenomena in Commerce. » Much of this section is devoted to constructing Hayek's (1941) model of capital structure by non-linear operations modeling verbally described there. ]

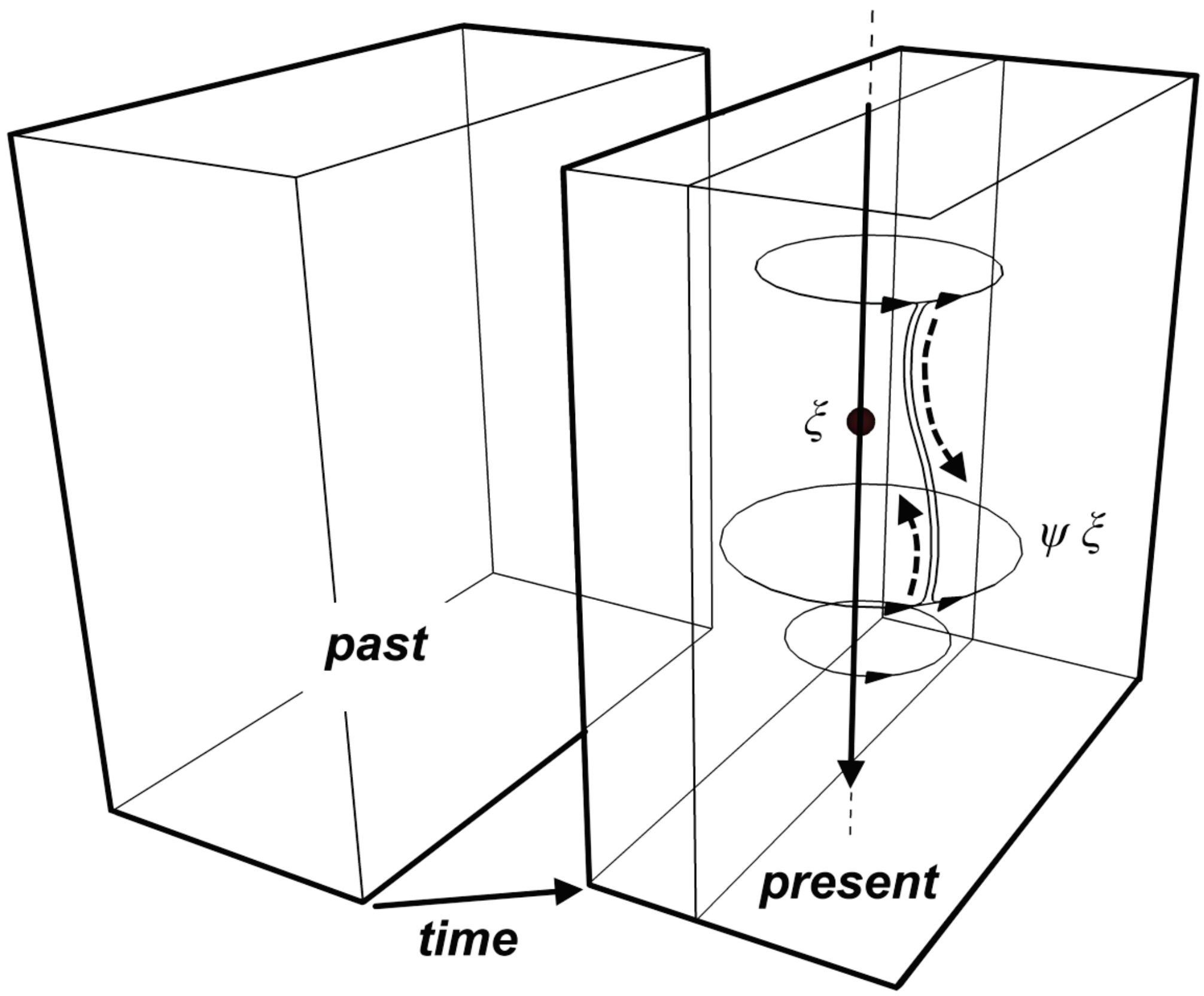

*Figure* 3. Following Imai's (1992) suggestion we can imagine $\psi\ \xi$ as complex velocity of series of various uniform distributions of sources and vortices which are spread around some path. Using Thomson's (1868) theorem we can visualize vortex layer $\psi\ \xi$ where complex flow of some infinitesimal element is identical everywhere along distinguishable circulations.

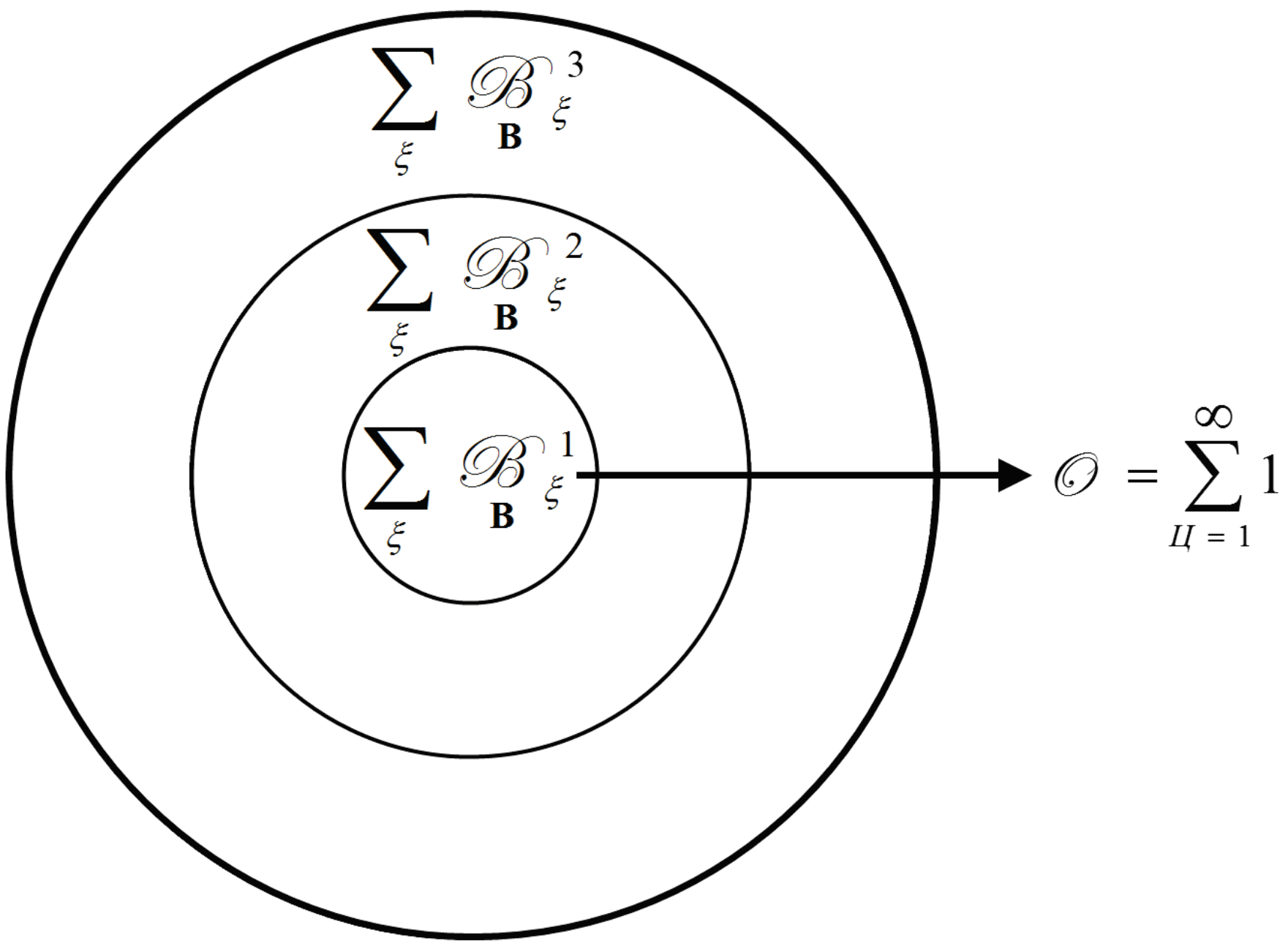

*Figure* 4. Bohm-Bawerk's (1889) structure of production diagrams are derived by summation of frequencies of different events according to their order in manufacturing processes during one concrete interval of time.

## §5

## Corollaries

[ Integration with a model of demand that is derived from the model of consumer preferences provided in: « Leibniz's Law of Continuous Change and Analysis of Fundamentally Uncertain Phenomena in Commerce. » ]

## §6

## Discussion

If marginal-productivity gains by way of division of labor in knowledge and work are observed from perspective of entrepreneurs or managers, « endo-physically », from perspective of parts of those systems whose behavior is studied (Rossler 1998), which is how all manufacturing processes are observed in actual practice by those who participating in them, then CBVP and KBVP are found to be alternate but complementary descriptions of identical underlying manufacturing processes. Neither CBVP is reducible to KBVP nor KBVP is reducible to CBVP.

Essentially there is no contradiction between knowledge determining both folding $C$ and map $\omega$ and so division of labor and structure of production and at the same time being a resource which also enters into production, with its own input-output coefficient $\psi_{\mu,K}$. It can assign itself a value based on results of this productive process in terms of revenue realized subtracting revenue foregone by that choice of particular manufacturing process.

Specifically, by determining the folding of manifold $\mathscr{M}$, it determines location of a bottleneck in production. This is located there where cross section $IK$ is minimal: one task is performed without sufficiently specialization and unless forbidden by a high market interest rate, it shall likely be found that marginal contribution of that task to revenue increases more than cost of performing it after additional capital is invested into more specialized performance of that task. This discovery is due to knowledge. What is marginal productivity of that piece of knowledge? This is necessarily difference in revenue associated with returns to this particular additional investment of capital at that point in the manufacturing process.

§7